\documentclass[prb,twocolumn,superscriptaddress,preprintnumbers,a4paper,amsmath,amssymb,showpacs,floatfix]{revtex4}
\usepackage{graphicx}
\usepackage{color}\color[rgb]{0.000,0.000,0.000} 
\usepackage{amssymb} 
\usepackage{amsmath} 
\usepackage{multirow}

\newcommand{\cu}{CuMnO$_{2}$}
\newcommand{\na}{NaMnO$_{2}$}

\begin{document}

\title{Decoupling lattice and magnetic instabilities in frustrated CuMnO$_{2}$}

\author{Keith V. Lawler}
\email{keith.lawler@unlv.edu}
\affiliation{Department of Chemistry and Biochemistry, University of Nevada Las Vegas, Las Vegas, Nevada 89154, USA.}

\author{Dean Smith}
\affiliation{Department of Physics and Astronomy and HiPSEC, University of Nevada Las Vegas, Las Vegas, Nevada 89154, USA.}

\author{Shaun R. Evans}
\affiliation{Formerly at: European Synchrotron Radiation Facility, Grenoble, France.}

\author{Antonio M. dos Santos}
\affiliation{Neutron Scattering Division, Oak Ridge National Laboratory, Oak Ridge, Tennessee, 37831, USA.}

\author{Jamie J. Molaison}
\affiliation{Neutron Scattering Division, Oak Ridge National Laboratory, Oak Ridge, Tennessee, 37831, USA.}

\author{Jan-Willem G. Bos}
\affiliation{Institute of Chemical Sciences, Centre for Advanced Energy Storage and Recovery, School of Engineering and Physical Sciences, Heriot-Watt University, Edinburgh, EH14 4AS.}

\author{Hannu Mutka}
\affiliation{Institut Laue-Langevin, 71 avenue des Martyrs, CS 20156, 38042 Grenoble, France.}

\author{Paul F. Henry}
\affiliation{ISIS Pulsed Neutron \& Muon Facility, Rutherford Appleton Laboratory, Harwell Campus, OX11 0QX, United Kingdom}

\author{Dimitri N. Argyriou}
\affiliation{European Spallation Source ERIC, PO Box 176, SE-221 00 Lund, Sweden.}

\author{Ashkan Salamat}
\email{salamat@physics.unlv.edu}
\affiliation{Department of Physics and Astronomy and HiPSEC, University of Nevada Las Vegas, Las Vegas, Nevada 89154, USA.}

\author{Simon A. J. Kimber}
\email{simon.kimber@u-bourgogne.fr}
\affiliation{Universit\'e Bourgogne-Franche Comt\'e, Universit\'e de Bourgogne, ICB-Laboratoire Interdisciplinaire Carnot de Bourgogne, B\^atiment Sciences Mirande, 9 Avenue Alain Savary, B-P. 47870, 21078 Dijon Cedex, France.}


\begin{abstract}
The $A$MnO$_{2}$~delafossites ($A$=Na, Cu), are model frustrated antiferromagnets, with triangular layers of Mn$^{3+}$~spins. At low temperatures ($T_{N}$=65 K), a $C2/m \rightarrow P\overline{1}$ transition is found in \cu, which breaks frustration and establishes magnetic order. In contrast to this clean transition, $A$=Na only shows short-range distortions at $T_N$. Here we report a systematic crystallographic, spectroscopic, and theoretical investigation of \cu. We show that, even in stoichiometric samples, non-zero anisotropic Cu displacements co-exist with magnetic order. Using X-ray/neutron diffraction and Raman scattering, we show that high pressures acts to decouple these degrees of freedom. This manifests as an isostuctural phase transition at $\sim$10 GPa, with a reversible collapse of the $c$-axis. This is shown to be the high pressure analog of the $c$-axis negative thermal expansion seen at ambient pressure. DFT simulations confirm that dynamical instabilities of the Cu$^{+}$~cations and edge-shared MnO$_{6}$~layers are intertwined at ambient pressure. However, high pressure selectively activates the former, before an eventual predicted re-emergence of magnetism at the highest pressures. Our results\footnote{This manuscript has been co-authored by UT-Battelle, LLC under Contract No. DE-AC05-00OR22725 with the U.S. Department of Energy. The United States Government retains and the publisher, by accepting the article for publication, acknowledges that the United States Government retains a non-exclusive, paid-up, irrevocable, worldwide license to publish or reproduce the published form of this manuscript, or allow others to do so,for United States Government purposes. The Department of Energy will provide public access to these results of federally sponsored research in accordance with the DOE Public Access Plan (http://energy.gov/downloads/doe-public-access-plan).} show that the lattice dynamics and local structure of \cu~are quantitatively different to non-magnetic Cu delafossites, and raise questions about the role of intrinsic inhomogeniety in frustrated antiferromagnets.

\end{abstract}
\maketitle
\section{Introduction}
Materials based upon the  $ABO_{2}$ delafossite structure have attracted much attention for their magnetic properties~\cite{marquardt2006crystal,goodenough1991lattice,onoda1993role,frontzek2011}.  This structure type consists of triangular layers of edge-sharing $BO_{6}$~octahedra, which are separated by layers of non-magnetic $A^{+}$~cations. Of particular recent interest have been the $A$MnO$_{2}$~materials with $A$= Na, Cu. This pair of materials are monoclinically distorted due to Mn$^{3+}$ orbital order, and differ only in the coordination of the interlayer cations. In the $A$=Na material, rocksalt coordination is found, whereas in \cu, linear O-Cu-O units are present (Fig. 1a). The differing nature of their magnetic ground states, and magnetoelastic couplings, has been much discussed. While both show rather 1D magnetism\cite{stock2009one,dally2018amplitude,kimber2020spin}, magnetic order causes broadening of the nuclear Bragg reflections in \na, implying an instability to a  triclinic structure \cite{giot2007magnetoelastic}. However, long-range symmetry breaking is not quite achieved, resulting in a nano-structured and inhomogeneous ground state\cite{zorko2014frustration}. In contrast, a bulk phase transition is seen for \cu, with a well-ordered triclinic ground state\cite{damay2009spin,vecchini2010magnetoelastic,terada2011magnetic} at the N\'{e}el temperature ($T_{N}$) of 65 K. This distortion lifts magnetic frustration, as the [110] in-plane couplings are no longer equivalent by symmetry. Curiously, neutron scattering\cite{frandsen2020nanoscale}~shows that short-range fluctuations of Mn-Mn distances persist well above $T_{N}$~for \na~only. The interlayer chemistry has further subtle effects in \cu. For example, the presence of antisite cation disorder changes the sign of the interlayer magnetic coupling from antiferromagnetic (AFM) to ferromagnetic (FM)\cite{garlea2011tuning,poienar2011substitution}. This changes the magnetic propagation vector (in the $P\overline{1}$ cell) from ($\frac{1}{2},\frac{1}{2},\frac{1}{2}$) to  ($\frac{1}{2},\frac{1}{2},0$).\\
Explanations for the different structural ground states of \na~and \cu~have focused on the role of the Cu vibrations, as well as postulated disorder. Vecchini $et~al$~used a stoichiometric sample and symmetry analysis to show that shear deformation of the MnO$_{6}$~layers  can be reinforced by the underbonded interlayer cations\cite{vecchini2010magnetoelastic}. In simple crystal chemical terms, this reflects the stronger covalent bonding along the Mn-O-Cu linkages, as compared to Mn-O-Na. This picture is also supported by theoretical estimates of the magnetoelastic coupling\cite{zorko2015magnetic}. In contrast, Frandsen $et~al$~used pair distribution function analysis to show that the instantaneous structure of \na~is actually more triclinic than \cu~on the local scale\cite{frandsen2020nanoscale}. This was said to indicate that residual disorder explains the different structural ground states, and that the $P\overline{1}$~structure reflects the non-disordered limit. Note that this work used a sample of \cu~which clearly displays both magnetic propagation vectors. In this work, we report detailed neutron and synchrotron X-ray scattering experiments for a sample of \cu~with no (measurable) anti-site disorder. We confirm the presence of highly anisotropic Cu displacements, which as in other delafossites, cause negative thermal expansion\cite{li2002strong,li2005trends, ahmed2009negative}~of the stacking direction ($c$-axis). However, in \cu, these are perpendicular to the strongest magnetic exchange interaction, and almost 1D in nature. We link these observations to an intense low-energy phonon mode, and show that substantial Cu positional disorder is found as T $\rightarrow$ 0 K, as parameterized by a residual anisotropic atomic displacement parameter. These observations are consistent with the previous symmetry analysis\cite{vecchini2010magnetoelastic}, as well as $^{63,65}$Cu NMR and NQR spectra for \cu, which show an interplay between magnetism and distortion of the local Cu$^{+}$~environment\cite{zorko2015magnetic}.\\
Given the highly anisotropic delafossite structure, the application of high pressure may decouple the Cu motions from the magneto elastic instability of the MnO$_{6}$~planes \cite{vecchini2010magnetoelastic}. Encouragingly, such experiments\cite{zhao1997x,xu2010pressure,salke2015raman} have uncovered rich behaviour in related materials like the arisotype delafossite CuFeO$_{2}$. Most strikingly, charge transfer\cite{xu2010pressure}~was found above  P = 23 GPa. This leads to a novel Cu$^{1+\delta}$/Fe$^{3-\delta}$~oxidation state in about 2/3 of the cations. Subsequent work showed that this mixed-phase state persists over a remarkably broad range of pressures\cite{xu2016cufe}. Inspired by these results, we also report an investigation of the lattice dynamics and structure of \cu~at pressures of up to 24 GPa. Our results show an isostructural phase transition at $\sim$10 GPa, which is driven by a softening of O-Cu-O vibrational modes. Using DFT simulations, we show how this couples to the lattice and magnetism, allowing a definitive separation of these instabilities. The O-Cu-O units of \cu, and their low energy dynamics, are thus confirmed to influence both the ambient and high pressure structures. Furthermore, this degree of freedom behaves quantitatively differently to non-magnetic Cu delafossites, where anisotropic atomic displacement factors extrapolate to zero at low temperatures \cite{li2002strong,li2005trends,ahmed2009negative}, and lattice hardening under pressure is found\cite{garg2018copper}.
\section{Experimental}
\textbf{Synthesis} Polycrystalline \cu \  was synthesised from CuO (Aldrich, 99.995 \%) and MnO (Aldrich, 99.99 \%). Stoichiometric quantities were mixed, pelleted and heated to 960 $\,^{\circ}\mathrm{C}$ in a sealed quartz ampoule under vacuum for 3 x 12 hours with intermediate regrinds. The heating rate was 5 $\,^{\circ}\mathrm{C}$ per minute, while cooling was achieved by switching off the furnace. The product was found to be a well-sintered blue-black solid, and phase purity was checked using laboratory X-ray diffraction.\\
\textbf{Ambient pressure crystallography} We performed neutron powder diffraction in the temperature range 2 $<$ T $<$ 300 K, using the E9 powder diffractometer\cite{tobbens2001e9}~at the former BER-II reactor of the Helmholtz-Zentrum Berlin. Temperature control was provided using an Orange He flow cryostat, and the sample was held in an 8 mm diameter vanadium can. At 2 K, we collected data using two settings of the Ge monochromator. These were 1.3084 \AA~using the (711) reflection, and 1.797 \AA~using the (511) reflection. Temperature dependent data were collected using the (511) reflection due to the higher thermal flux at this incident wavelength. Some additional data points were also collected on the D20 diffractometer at the Institut Laue Langevin. These used the 1.87 \AA, high take off angle (120$^{\circ}$) configuration and covered the range 6-75 K.  Crystal struture models were refined using the Rietveld method as implemented in the GSAS-II software\cite{toby2013gsas}.\\
We also collected synchrotron X-ray scattering data, suitable for transformation into pair distribution functions (PDFs). These experiments were made on the former ID15B station of the European Synchrotron Radiation Facility, using an incident energy of 87.3 keV. The sample was placed in a borosilicate capillary, and spun during data collection. Temperature control used an Oxford cryostream, and the scattered X-rays were collected using a Mar345 image plate. The raw images were azimuthally integrated using pyFAI\cite{ashiotis2015fast}, and transformed into the PDFs using PDFGetX3 and an experimentally determined background\cite{juhas2013pdfgetx3}. The Q-max value used varied between 26.8 \AA$^{-1}$~at low temperature, to 25.5 \AA$^{-1}$~at room temperature. We refined crystal structure models using PDFGui in the small-box approximation\cite{farrow2007pdffit2}.\\
\textbf{Inelastic neutron scattering} Time of flight inelastic neutron scattering data were collected using the IN4 instrument at the high flux reactor of the Institut Laue-Langevin, Grenoble. A large ($>$5 g) sample was placed in a rectangular Aluminium foil envelope which was then clamped in a Cd frame and attached to the cryostat stick. Data were collected at 150 K at an incident energy of 67 meV. The data were normalised to vanadium and a background subtracted using standard routines in the \textsc{LAMP}~software. Cuts through the inelastic response were parameterised using the damped harmonic oscillator model given below. This was implemented in Igor Pro, and was convoluted with the instrumental resolution determined by fits to the elastic line of the vanadium data. The model\cite{faak1997phonon,Loidl}~includes the oscillator frequency ($\omega_{DHO}$), line width ($\Gamma_{DHO}$) and the final term accounts for detailed balance at the experimental temperature, $T$.
\begin{equation}
S(Q,\omega)=\frac{A_{\textrm{DHO}}\omega\Gamma_{\textrm{DHO}}}{(\omega^2-\omega^2_{\textrm{DHO}})+(\omega\Gamma_{\textrm{DHO}})^2}\cdot\frac{1}{1-e^{-\hbar\omega/k_BT}}
\end{equation}\\
\textbf{High pressure Raman} Raman spectroscopy was performed on a home-built system employing OptiGrate spectral filters, a $f$/9 Princeton spectrometer, and excitation with the 488\,nm line of an argon-ion laser not exceeding 5\,mW.
Measurements were taken on pressure increase to 20\,GPa on \cu{} samples in He pressure transmitting medium in a membrane-driven diamond anvil cell (DAC) with ruby photoluminescence to calibrate pressure.\\
\noindent\textbf{High pressure XRD} For synchrotron powder diffraction experiments at high pressure, a finely ground sample was gas loaded (He) in a membrane driven diamond anvil cell. A spring steel gasket was used, and the pressure calibrated using the fluorescence of a small chip of ruby placed in the cell. Data were collected on the former ID09A station at the ESRF (operations now moved to ID15B), using a Mar555 detector. The sample-detector distance was calibrated using NIST Si, and the data were azimuthally integrated using Fit2D\cite{hammersley2016fit2d}. The incident x-ray energy was 30 KeV, and the maximum pressure reached was 24 GPa. Data were collected on compression and decompression. Structural refinements were performed using \textsc{GSAS}~and the \textsc{EXPGUI}~interface\citealp{larson1994gsas,toby2001expgui}.\\
\textbf{High pressure neutron diffraction} 
Neutron powder diffraction experiments were performed using the SNAP time of flight diffractometer at the Spallation Neutron Source, ORNL\cite{calder2018suite}. The center wavelength was set at 2.1 \AA, yielding an available wavelength range from 0.5 \AA~to 3.5\AA. This instrument is fitted with two independent detectors, placed right and left of sample. These cover +/- 22.5$^{o}$ in angular range in and out of plane and were centered at nominally 75$^{o}$ and 103$^{o}$. The combination of wavelength range and angular spread, allowed the collection of data from 0.5 $-$ 3~\AA, an appropriate range for the small unit cell of CuMnO$_{2}$~. The sample was pressed into a dense sphere and contained in a null scattering alloy, TiZr,  gasket. Pressure was applied using a Paris-Edinburgh press \cite{besson1995new}~equipped with polycrystalline diamond anvils with a double toroidal geometry \cite{Khvostantsev2004_DT_Anvils, Fang2012_DT_Anvils}. We reached a maximum pressure of $\sim$17 GPa, however the sample signal deteriorated strongly above 12 GPa. A novel refinement strategy was employed with the sample used as the pressure calibrant (described in the text in more details). The TOPAS software was used to perform these refinements\cite{coelho2018topas}.\\
\noindent\textbf{Simulations}
Plane-wave density functional theory (DFT)\cite{DFT-HK,DFT-KS} simulations were performed with VASP 5.4.4. 
The SCAN\cite{SCAN-DFT} meta-GGA functional was employed with a 600 eV plane-wave cutoff and 10$^{-8}$ eV energy tolerance. The 3d states were treated with a single parameter Hubbard DFT+U\cite{Dudarev_LSDAU_1998,Wang_TMO+U_2006} correction, 3.0 eV for Mn and 7.0 eV for Cu. 
A $\Gamma$-centered Monkhorst-Pack\cite{MPgrid} grid with 0.05$\times$2$\pi$\AA$^{-1}$ spacing represented the first Brillouin zone.
All structural relaxations minimized forces below 10$^{-3}$ eV/\AA{}. 
Phonons were calculated at the $\Gamma$-point by central difference finite displacement (0.015\,\AA). 
Localized orbital bonding analysis (LOBA)\cite{LOBA} of molecular sub-units was performed with Q-Chem 5.0\cite{B915364K,doi:10.1080/00268976.2014.952696} using the PBE\cite{PBE} functional, the def2-TZVP\cite{B508541A} basis set, and a 10$^{-9}$ E$_h$ energy tolerance.

\section{Results}
\subsection{Ambient pressure characterisation}
The phase purity and structure of our sample was determined by the Rietveld method and the room temperature neutron powder diffraction data. An excellent fit was achieved using the $C2/m$~structure, and a small (2.6 \%~weight fraction) cubic spinel impurity. Site mixing between Cu and Mn has been reported elsewhere\cite{garlea2011tuning,poienar2011substitution,terada2011magnetic}, and we tested this for our sample by constraining and refining the site occupancies appropriately. However, no statistically significant deviation from complete occupation was found, despite the excellent contrast between scattering lengths ($b_{Mn}$=-3.73 and $b_{Cu}$=7.718 fm). As an aside, we note that direct substitution of an octahedral Mn species for Cu$^{+}$~is unlikely due to the highly different bonding. In our view, more work needs to be done to characterise any local relaxation around such defects. The best fits were achieved when we allowed for an anisotropic displacement parameter on the Cu site\cite{li2002strong}. The resulting ellipsoids (100 \% level) are shown\cite{momma2011vesta}~in Fig. 1a, and can be seen to be dominated by motions transverse to the $c$-axis, as expected for linear O-Cu-O coordination. 
\begin{figure}[]
  \includegraphics[width=7.5 cm]{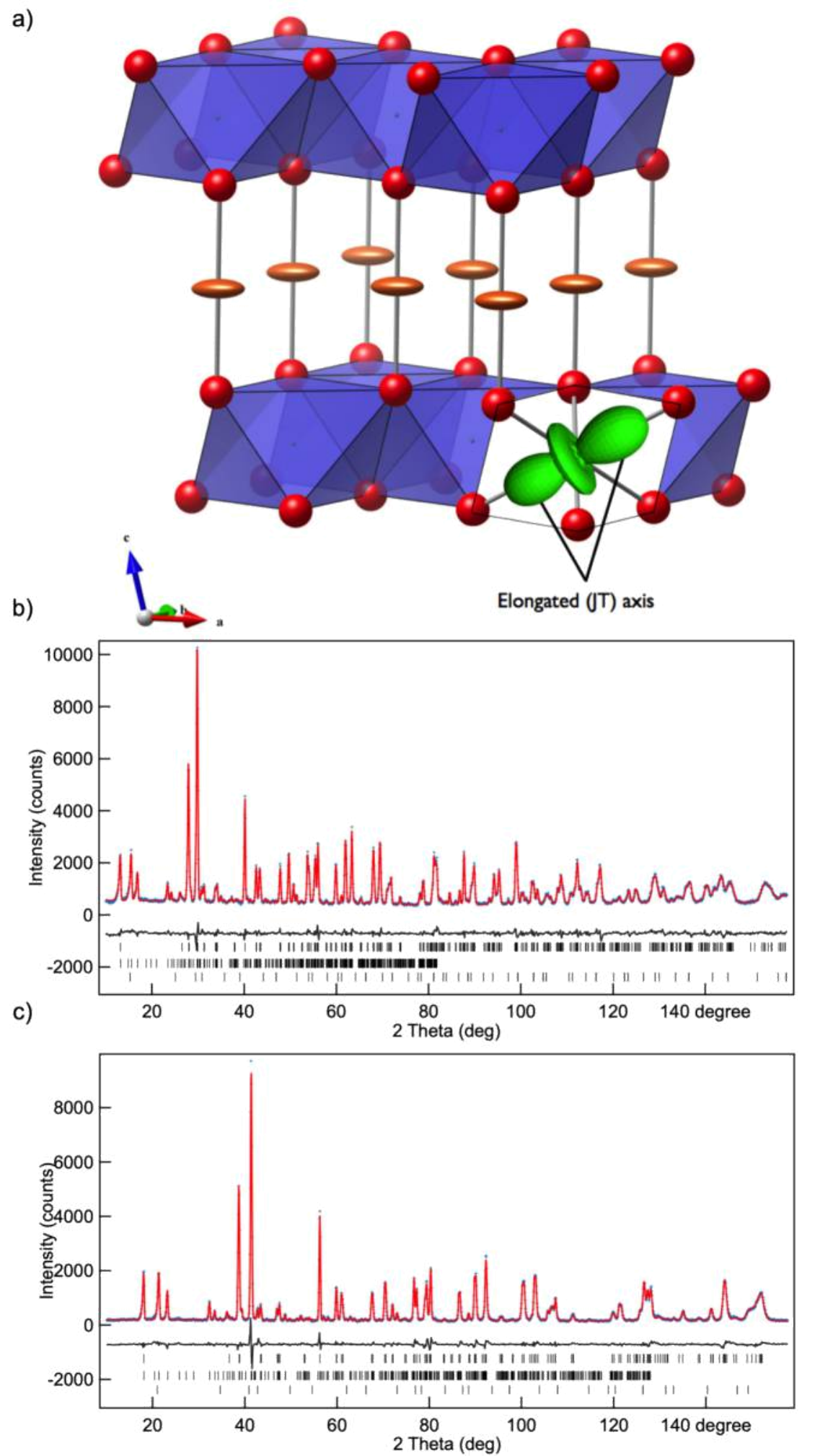}
  \caption{a) Crystal structure of monoclinic \cu, showing the MnO$_{6}$~octahedra in blue, and the linearly coordinated Cu$^{+}$~cations in bronze. The latter show the anisotropic thermal ellipsoids extracted from neutron powder diffraction data at 300 K. One MnO$_{6}$~octahedra is shown in cutaway, with the preferentially occupied $d_{z^{2}}$~orbital responsible for the Jahn-Teller distortion indicated; b) Observed, calculated and difference profiles for the combined fit to the 2 K high resolution neutron powder diffraction data collected using 1.797 \AA~neutrons; c) Observed, calculated and difference profiles for the combined fit to the 2 K high resolution neutron powder diffraction data collected using 1.308 \AA~neutrons. Three phases are shown (from top), triclinic $C\overline{1}$~\cu, the $k$=($\frac{1}{2},\frac{1}{2},\frac{1}{2}$) magnetic order, and a 2.6 \%~weight cubic spinel impurity. The refinement converged with an overall goodness-of-fit, GOF = 1.86.}
  \label{fgr:structure}
\end{figure}
Our sample also cleanly underwent the $C2/m$~to $P\overline{1}$~phase transition at 65 K. We performed a precise refinement of the latter structure at 2 K (Fig. 1 b) by co-refining two data sets with different wavelengths. As described later in more detail, this confirmed the presence of anisotropic Cu displacements even at 2 K. The value of the $U_{11}$~component of the ADP tensor refined to  0.0065(5) \AA$^{2}$. Other components of the tensor refined to very small values, including some which have small negative values. This reflects the presence of a small absorption effect. We have avoided applying an \textit{ad hoc} correction, since this simply adds an arbitrary offset. Note that the ADP tensor is positive definite at all higher temperatures.  All of the 'cross' ADP parameters ($U_{12}$, $U_{13}$, $U_{23}$) were zero, and we fixed these to zero in the $C\overline{1}$~cell used for refining the triclinic structure as a function of temperature. Finally, note that this refinement also included the $\vec{\textbf{k}}=(-\frac{1}{2},\frac{1}{2},\frac{1}{2}$) magnetic structure expected\cite{garlea2011tuning}~for stoichiometric \cu. Full details of the refined structure may be found in table I\\
\begin{table}[h!]
  \begin{center}
    \caption{Details of the refined crystal structure of CuMnO$_{2}$ at 2 K, using the two wavelength neutron powder diffraction approach described in the main text. Refinements were performed using a $C\overline{1}$ unit cell for ease of comparison to the monoclinic structure. As described in the text, the slightly negative value of some thermal parameters reflects a small absorption effect. All cross terms in the anisotropic displacment factor tensor for Cu were zero within error.}
    \label{tab:table1}
    \begin{tabular}{c|c} 
      Parameter & 2 K \\
      \hline
      a (\AA) & 5.5754(5) \\
      b (\AA) & 2.88017(2) \\
      c (\AA) & 5.8932(5) \\
      $\alpha$ ($^\circ$) & 90.172(2) \\
      $\beta$ ($^\circ$) & 103.937(2) \\
      $\gamma$ ($^\circ$) & 89.827(2) \\
    O(x) & 0.40650(19) \\ 
    O(y) & -0.0004(6) \\ 
    O(z) & 0.17824(15) \\ 
    U$_{Mn}$ (\AA$^2$) & -0.0008(3)\\ 
    U$_{O}$ (\AA$^2$) & 0.00220(26) \\ 
    U$_{11}$ Cu (\AA$^2$) &  0.0065(5)\\ 
    U$_{22}$ Cu (\AA$^2$) & -0.0012(4) \\ 
    U$_{33}$ Cu (\AA$^2$) & -0.0021(4) \\ 
    \end{tabular}
  \end{center}
\end{table}

The temperature dependence of the Mn-Mn distances is shown in Fig. 2a, highlighting the symmetry breaking in the plane, as well as the pronounced negative thermal expansion along the $c$-axis. This goes through a minimum around 200 K. We also show the temperature dependence of selected Cu ADP components (Fig. 2b). The temperature dependent data can be seen to be consistent with both our precise results at 2 K, as well as literature\cite{damay2009spin}~values reported at 300 K.
\begin{figure}[]
  \includegraphics[width=7 cm]{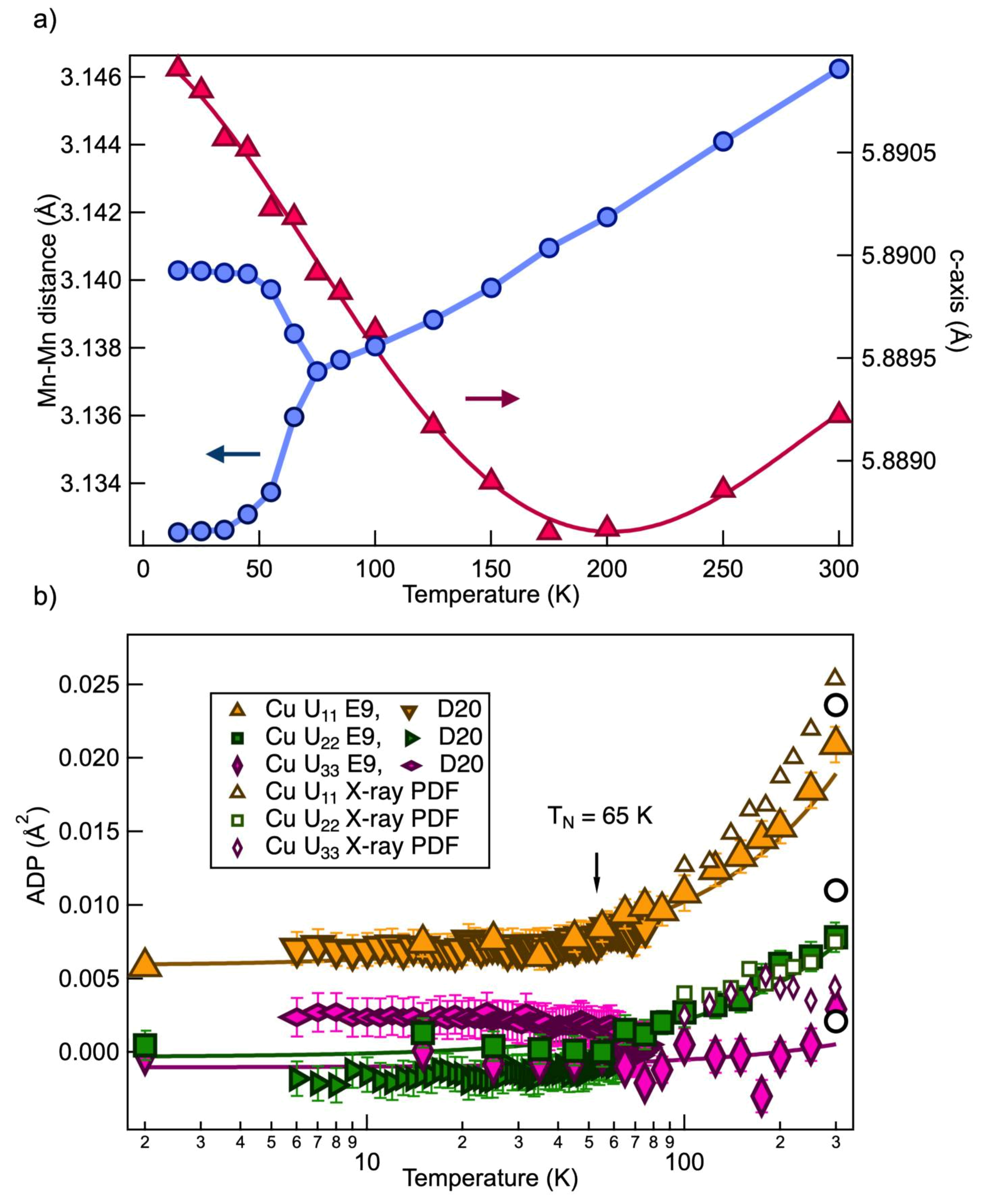}
  \caption{a) Temperature dependence of the Mn-Mn distances extracted from our E9 data, showing the symmetry breaking at $T_{N}$=65 K; b) Temperature dependence of the anisotropic atomic displacement parameters for Cu, as extracted from neutron powder diffraction and X-ray pair distribution function measurements. Note the large residual value of the $U_{11}$~parameter. Open black circles show the values reported\cite{damay2009spin}~by Damay \textit{et al}~at 300 K. Lines are guides to the eye}
  \label{fgr:structure}
\end{figure}
Additional data points from independent refinements against our D20 data also follow the same trends. Surprisingly, and at odds with other delafossites\cite{li2002strong,li2005trends,ahmed2009negative}, the Cu motion is highly 1D. The refined value of the other in-plane component of motion ($U_{22}$) is small, and comparable to the Mn$^{3+}$~and O$^{2-}$~isotropic displacement parameters at all temperatures. As expected, the $U_{33}$~component is negligible , reflecting the highly anisotropic bonding. Using a simplified Einstein oscillator model\cite{sales1997filled}~and the E9 data, we extracted an energy scale for vibrations along the [100] direction. This analysis gives an estimate of 10.8 meV or 87 cm$^{-1}$ for the frequency of oscillation.\\
\begin{figure}[]
  \includegraphics[width=8 cm]{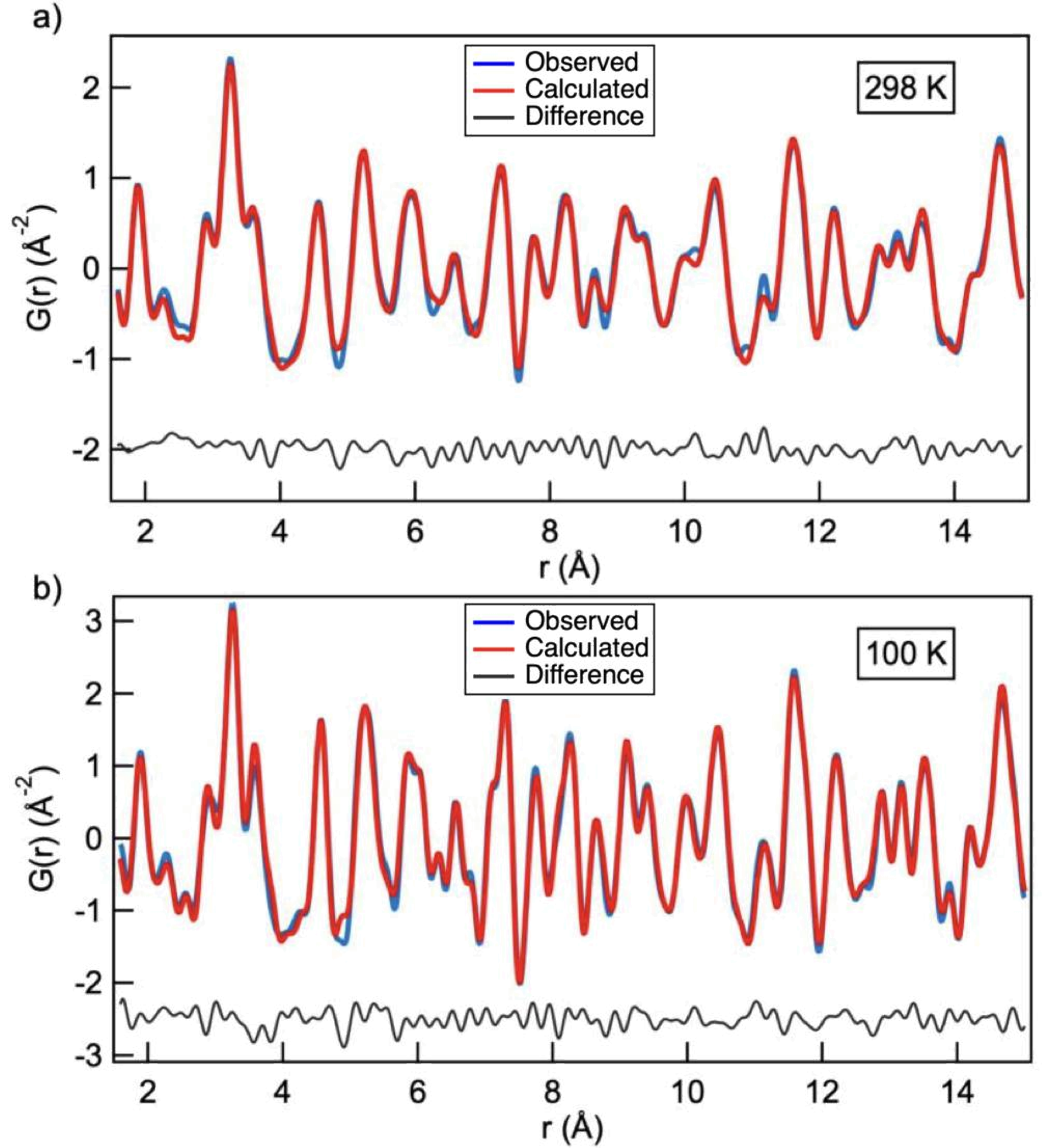}
  \caption{a) Observed, calculated and difference profiles for the fit of the $C2/m$~structure to the X-ray pair distribution function of \cu~at 298 K. This refinement converged with $R_{w}$=0.127; b) Observed, calculated and difference profiles for the fit of the $C2/m$~structure to the X-ray pair distribution function of \cu~at  100 K. This refinement converged with $R_{w}$=0.125.}
  \label{fgr:structure}
\end{figure}
Examination of our pair distribution functions (Fig. 3), showed no major changes between 80 and 350 K. This rules out any short-range triclinic domains, in agreement with recent work\cite{frandsen2020nanoscale}. We therefore proceeded to refine the monoclinic $C2/m$ ~model as a function of temperature over the range 1.6 $<$~r~$<$15 \AA. We achieved an equally good fit of this structure at 100 and 298 K. However, much as in the neutron powder diffraction experiments, this required anisotropic Cu ADPs. The refined values of $U_{11}, U_{22}$~and $U_{33}$~were comparable to those extracted by Rietveld refinement, and showed the same temperature dependence (Fig. 2b). In contrast however, the PDF refinements always gave a non-negligible value of $U_{13}$. This has the effect of rotating the thermal ellipsoid around the $b$-axis, such that it is approximately parallel to the long Jahn-Teller distorted Mn-O axis. The origins of this effect are unclear, and will require future study. Finally, we also tried various split Cu site models, with Cu displaced along [010]. These gave equivalent fits to the PDFs.\\
We conclude this section by describing the powder inelastic neutron scattering response of \cu. At 150 K, this is dominated by two features. The first\cite{terada2011magnetic}~is a low-Q plume of magnetic scattering (which is characterised in more detail elsewhere\cite{kimber2020spin}), and the second is an intense and dispersion-less mode at ca. 11 meV, which increases in intensity as a function of momentum transfer, $Q=4\pi\sin(\theta / \lambda)$. 
\begin{figure}[]
  \includegraphics[width=8 cm]{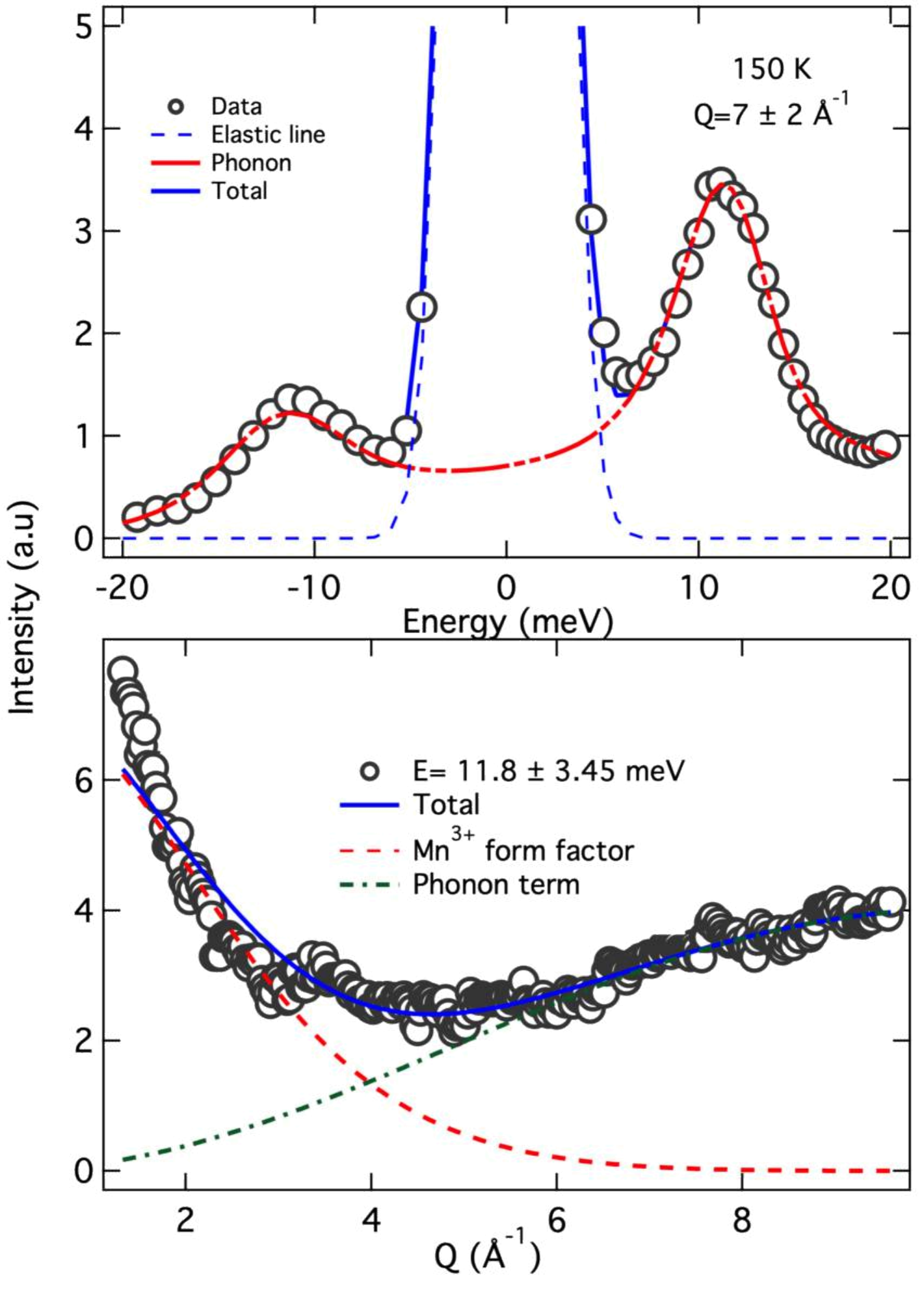}
  \caption{(top) Q-slice through the inelastic neutron scattering response of \cu~collected on IN4 at 150 K. Data are summed over the range Q=7$\pm$2 \AA $^{-1}$. The data are fitted with the damped harmonic oscillator model described in the text, which is convoluted with the instrumental resolution function and accounts for detailed balance; (bottom) Constant energy cut (E=11.8$\pm$~3.45 meV) through the data. The solid line is a sum of the Mn$^{3+}$~form factor squared and a $Q^2$~phonon term. This confirms the vibrational origin of the high-Q signal.}
  \label{fgr:IN4}
\end{figure}
Constant Q-cuts were parameterised using the damped harmonic oscillator model described above (Fig. 4b), while the vibronic nature of this feature was confirmed by constant energy cuts (Fig. 4a), which show a cross-over between a magnetic form-factor like response at low-$Q$ to $\propto Q^{2}$~behaviour at high-$Q$. The final fitted oscillator frequency was 11.9(2) meV, or $\sim$~96 cm$^{-1}$. 
\subsection{High pressure Raman scattering}
The arisotype trigonal/hexagonal delafossite structure has only two Raman allowed\cite{Aktas_2011} vibrations, which are normally described as having $e_{g}$~and $a_{1g}$~symmetry. These can be visualised in real space as in-plane vibrations of the triangular MO$_{6}$~layers and $c$-axis vibrations of the A-cation dumbell units respectively. The latter mode is typically the most intense, and is found between ca. 690 and 760 cm$^{-1}$~for a range of Cu containing delafossites\cite{Aktas_2011,Ramancuga,Ramancual}. The in-plane mode typically appears at lower energy around 350 - 420 cm$^{-1}$.
The ambient pressure Raman spectra of our sample of \cu~is shown in Fig. 5 (top), and appears to be consistent with other reports\cite{chen2015crednerite}. Using a Lorentzian line shape, we extracted the position of the intense mode at 688.70(3) cm$^{-1}$, which we assign to the $A_{g}$~vibration. The two weaker modes, which appear where the arisotype's $e_{g}$~mode is expected, are at 312.5(2) and 388.8(2) cm$^{-1}$ and we label these as $A_{g}^{IP}$~and $B_{g}^{IP}$~($vide~infra$). We were able to follow the pressure dependence of all three modes to 20 GPa, and their detailed evolution is shown in Fig. 5 (bottom). The $a_{1g}$~vibration hardens substantially up to 10 GPa before softening to well below its room temperature frequency. This structural change is also seen clearly in the IP1 mode, which hardens substantially, and weakly in the IP2 mode, which curiously softens immediately from ambient pressure. 
By using the linear region below 8 GPa, we extracted pressure coefficients of 3.9(2) ($A^{IP}_g$),  -0.718(2) ($B^{IP}_g$) and 4.2(2) cm$^{-1}$/GPa ($A_{g}$).
\begin{figure}[]
	\includegraphics[width=9 cm]{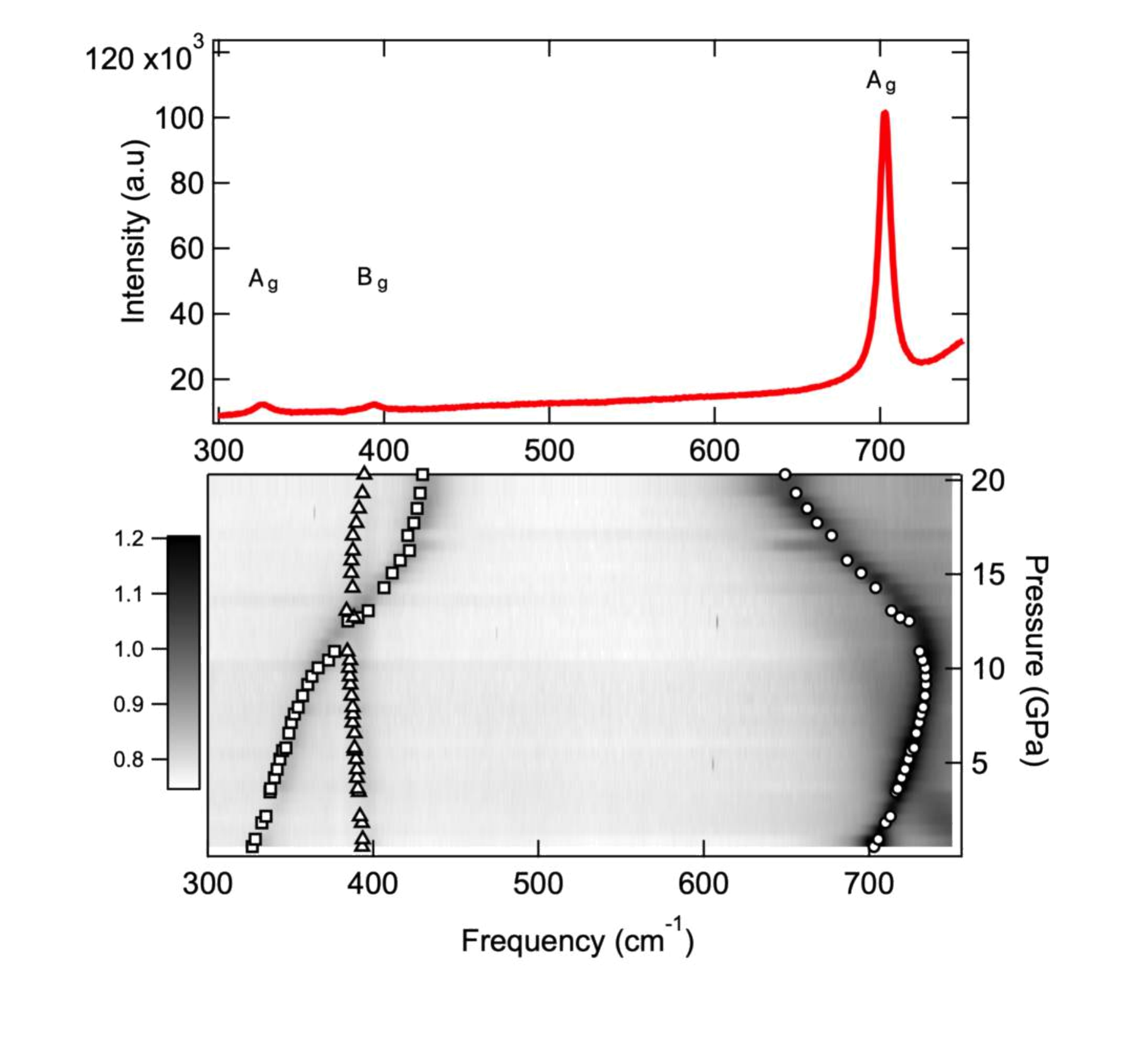}
	\caption{(top) Lowest pressure Raman spectra for \cu~obtained in the diamond anvil cell. (bottom) Intensity colour map and fitted peak positions for the pressure dependent Raman data. The intensity data have been normalised and a linear background removed. }
	\label{fgr:Raman}
\end{figure}
\subsection{High pressure x-ray diffraction}
The azimuthally-integrated data from ID09A were free from any contamination from \textit{e.g.} gasket \textit{etc.}, and the data sets at low pressure were indexed using the known $C2/m$~cell.
Unfortunately, the data were strongly influenced by heterogeneous crystallinity, with the apparent presence of many small crystallites with varying orientation.
This was despite grinding the sample well before loading, and was particularly pronounced due to the microfocusing needed to avoid the gasket when performing diamond anvil cell experiments (the beam size on ID09A was $\sim$30\,$\mu$m in diameter).
We therefore used Le Bail refinements to follow the pressure dependence of the lattice parameters only. As can be seen in Fig.~\ref{fgr:XRD}a, a change in the diffraction patterns is seen around 10 GPa, where a non-linear change in peak positions sets in.
The number of peaks remains the same throughout the range 0.5\,$<$\,P\,$<$\,24 GPa, and neighbouring peaks smoothly cross over in position.
These observations are consistent with an isostructural phase transition.
\begin{figure}[]
	\includegraphics[width=9 cm]{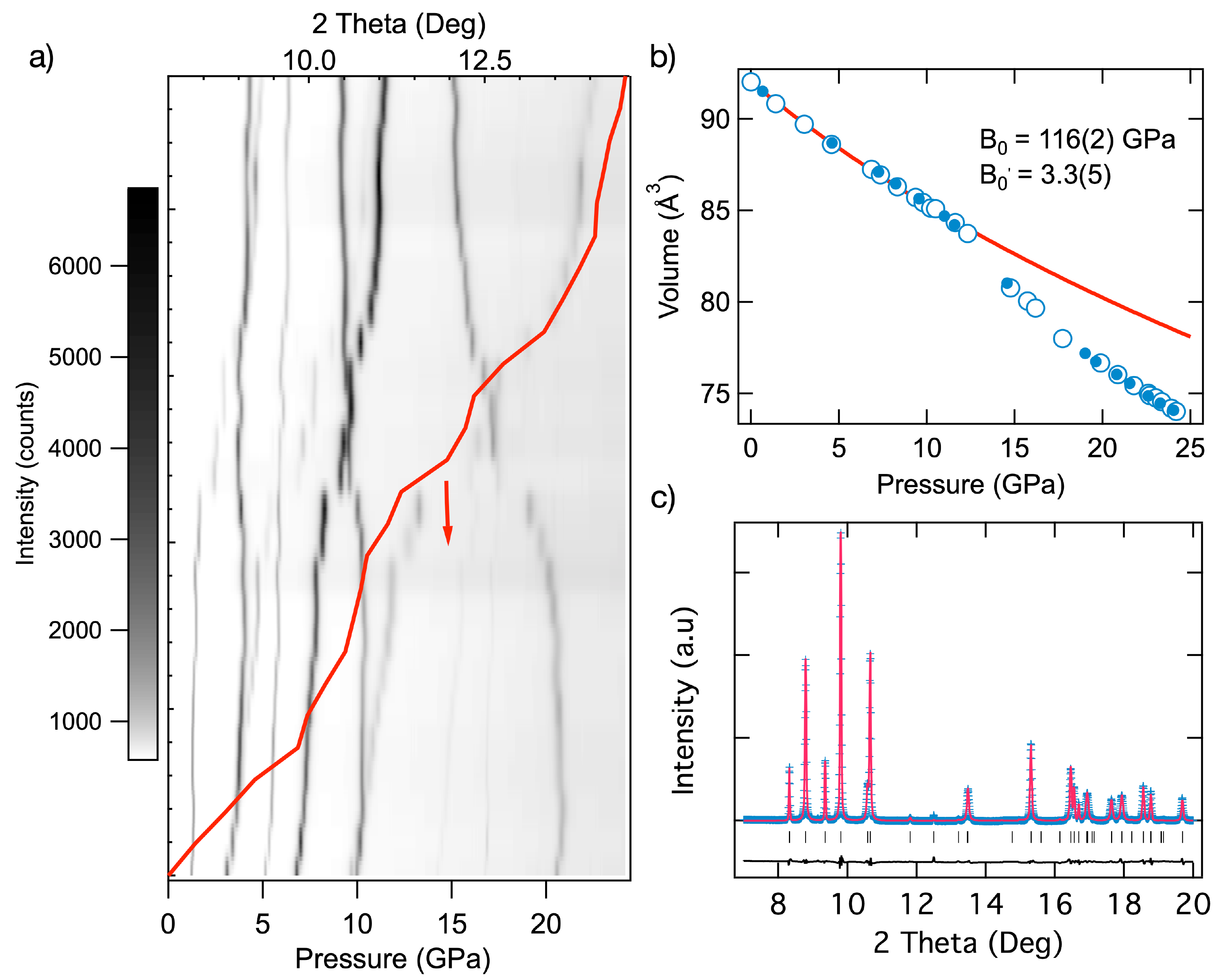}
	\caption{a) Stack plot of the x-ray diffraction data for \cu~collected at ID09A. Note that the pressure is increasing vertically, although the data points are not equidistant. The detailed evolution of the pressure is indicated by the solid red line and the bottom axis; b) Refined volume for \cu~as a function of pressure. The solid red line shows the fit of a third order Birch-Murnaghan equation of state to the region 0 $<$ P $<$ 10 GPa; c) Typical observed, calculated and difference profiles for a Le Bail fit of the monoclinic $C2/m$~structure. The data shown are at $\sim$2.5 GPa, and have been background subtracted.}
	\label{fgr:XRD}
\end{figure}
\begin{figure}[]
  \includegraphics[width=8 cm]{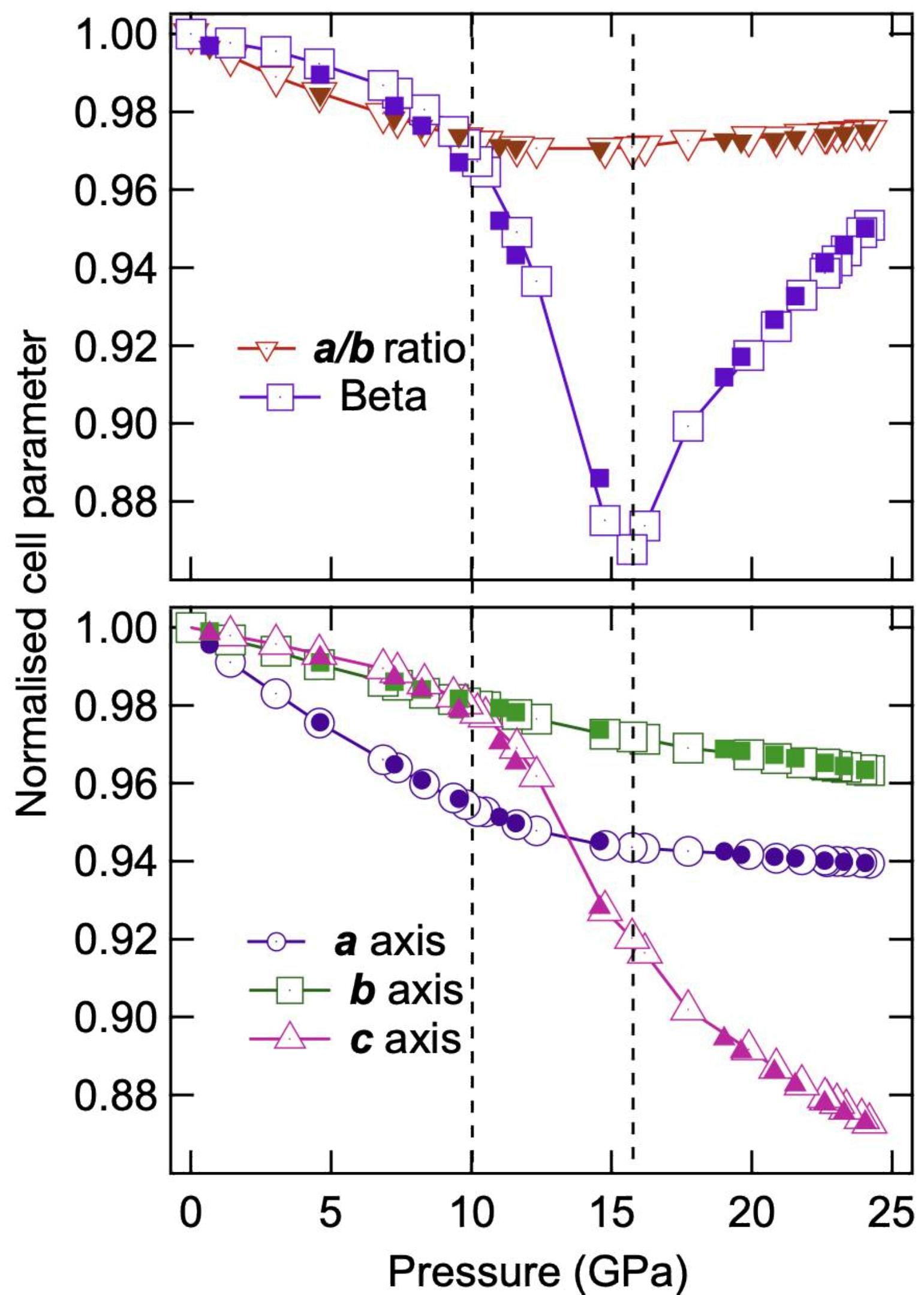}
  \caption{(top) Refined $a$/$b$~ratio and monoclinic angle for \cu~as a function of pressure. Note that these are shown normalised to the ambient pressure value, and are hence dimensionless.  The former is a measure of the distortion away from hexagonal symmetry in the MnO$_{6}$~layers. (bottom) Refined lattice parameters for \cu~as a function of pressure. The lines in both plots are guides to the eye, and all parameters have been normalised by their zero pressure values. Open/filled symbols represent data collected on compression/decompression respectively.}
  \label{fgr:lattice-params}
\end{figure}
Indeed, our efforts to index the high pressure data gave results equivalent to the  $C2/m$~cell. Performing refinements of this cell against the data gave the unit cell volume shown in Fig. 6b, with a clear volume anomaly around 10\,GPa. We were able to fit a 3$^\text{rd}$ order Birch-Murnaghan equation of state to the low pressure region. This yielded a bulk modulus, $B_{0}$=116(2)\,GPa with a pressure derivative, $B'$=3.3(5). As shown in Fig.~\ref{fgr:lattice-params}, the pressure response of the of the cell axes are highly anisotropic, both above and below the transition. We extracted compressibilities ($\beta$=-1/a($d$a/$d$P) etc) in the approximately linear regions below 8 GPa, yielding $\beta_{a}$=0.0048(2), $\beta_{b}$=0.00200(1), $\beta_{c}$=0.00156(3) GPa$^{-1}$.  However, upon exceeding 10 GPa, a switch is found, with the in-plane $a$~-axis stiffening, and a dramatic softening of the $c$~-axis (stacking direction). The monoclinic angle actually passes through 90 $^{\circ}$~just above 15 GPa, and trial Le Bail refinments in orthorhombic space groups showed the lattice to be metrically orthorhombic. However, the lack of any anomaly in the cell parameters/volume means that this is likely just an accidental pseudo-symmetry. Note that the phase transition is completely reversible, and in both Fig. 6b and Fig. 7, open symbols represent results obtained on increasing pressure, and filled points represent results obtained on decreasing pressure.
\subsection{High pressure neutron diffraction}
Given the problems encountered with the high pressure x-ray experiment, which arise from the combination of a highly crystalline sample and microfocused beam, we pursued high pressure neutron powder diffraction. This is highly complimentary, due to its sensitivity to oxide displacements. However, the angular resolution available is much lower that in the X-ray case. We therefore exploited advances in parametric Rietveld approaches\cite{Stinton}, as implemented in the TOPAS software to design our experiment. Since only the external load on the cell is controlled and not the samples pressure, one would usually include a pressure marker (e.g. Pb or NaCl) with a well known equation of state in the sample space. The pressure can then be back-calculated from the unit cell volume of the pressure marker. Since the X-ray diffraction experiment gave a very detailed pressure evolution for each of the unit cell axes, we decided to use the sample itself to calibrate the pressure. Polynomials were fitted to the data shown in Fig. 7, and instead of refining the unit cell parameters for each data set, we simply refined the pressure. A multiplicative offset term, which refined to 1.004(3) was included in case there were calibration shifts between ID09A and SNAP. Two loadings were used, the first of which reached an inlet-gas pressure of 700 bar before failing. The second loading was commenced with a inlet pressure of 650 bar in order to give some overlapping data points.
The Lorentzian and Gaussian contributions to the peak shapes and the sample offsets were refined globally, and constrained to change linearly with pressure. The moderator contributions to the peak shape were fixed to be the same for all histograms, and refined at the end of the procedure, resulting in small shifts from the values established with a nickel standard.\\
The contribution from the polycrystalline diamond anvils, and their tungsten carbide binders were refined using the Pawley approach \cite{pawley1981unit}. The lattice parameters of the WC were allowed to vary linearly with pressure, which is a reasonable approximation to the low-pressure behaviour of this extremely hard material. An independent sample displacement term was necessary for each of the two loadings, reflecting the difficulty of repositioning a $\sim$~30 kg pressure cell precisely. The only parameters which were freely refined at each pressure were therefore the two oxygen fractional coordinates, $O_{x}$~and $O_{z}$, and three isotropic atomic displacement parameters.
\begin{figure}[]
	\includegraphics[width=8 cm]{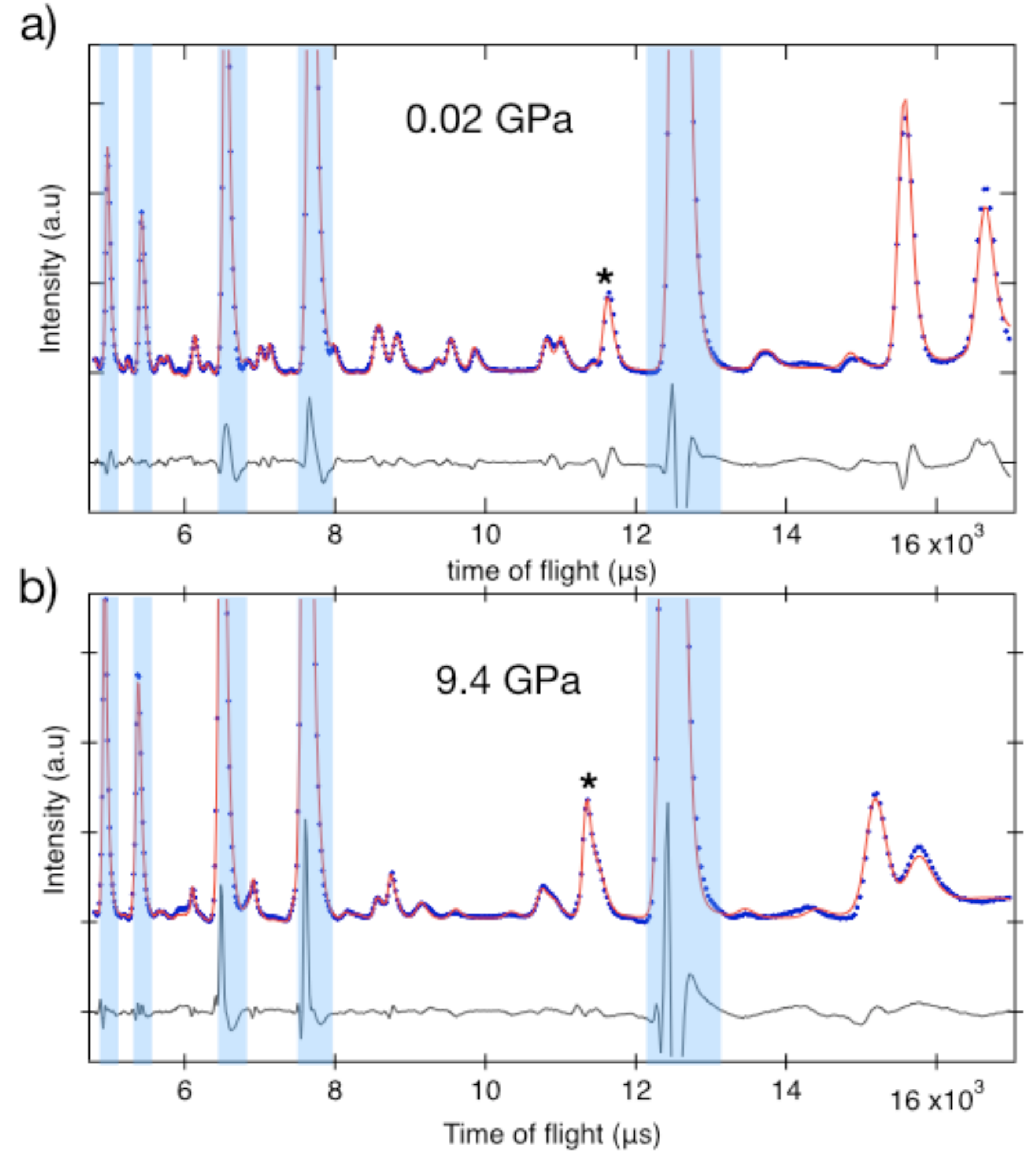}
	\caption{Observed, calculated and difference profiles from the Rietveld fits to the neutron powder diffraction data at 0.02 (top) and 9.4 (bottom) GPa. A prominent peak from the WC anvil binder is highlighted by a star, and the shaded regions contain the very intense reflections from the diamond anvils.}
	\label{fgr:XRD}
\end{figure}
Example Rietveld fits in the low and high pressure regions are shown in Fig. 8. Discounting the very intense reflections from the diamond anvils, the agreement is generally good, and we were able to freely refine the pressure for each data set and extract bond lengths. The load-pressure curves obtained are shown in Fig. 9a, where both can be seen to be approximately linear. As shown below, the small differences in pressure offset and slope between loadings do not produce any anomalies in the refined parameters at the cross over pressure of $\sim$~8 GPa. The pressure dependence of the two inequivalent Mn-O bonds is shown in Fig. 9b, where a marked difference in compressibility is seen between the long ($\beta$=0.0049(4) GPa$^{-1}$) and short ($\beta$=0.0008(1) GPa$^{-1}$) directions. Notably, the Jahn-Teller distortion, which can be parameterised using the individual ($l_i$) and average ($l_av$) bond lengths as:
\begin{equation}
D=\frac{1}{n}\sum^{n}_{i=1}\frac{\lvert l_{i}-l_{av}\rvert}{l_{av}}
\end{equation}
shows a weak pressure dependence, falling slowly from 7 x 10$^{-2}$~at 0 GPa to 5 x 10$^{-2}$~at 11 GPa. This is always well within the range considered characteristic\cite{Kimber_2012} for Jahn-Teller distorted Mn$^{3+}$, and appears to rule out charge transfer as a mechanism for the pressure induced collapse. Either oxidation or reduction of Mn$^{3+}$~yields non-JT active species, with distortion parameters in the (approximate) range 0 $<D<2$~x 10$^{-2}$. Curiously, a slight uptick in both bond lengths is seen above the transition pressure. This will be discussed in more detail below.
The compressibility of the O-Cu-O dumbell bonds, as well as the longer range Cu-O contacts were also extracted (Figs 9c and 9d). Below $\sim$~10 GPa, the former were as rigid as the short Mn-O distances, with $\beta$=0.0008(1) GPa$^{-1}$. In contrast, the longer non-bonded distances are much softer ($\beta$=0.0059(3) GPa$^{-1}$.). Dramatic softening of both is found at 10 GPa, suggesting a role for Cu-O contacts in the anisotropic collapse.
\begin{figure}[]
	\includegraphics[width=9 cm]{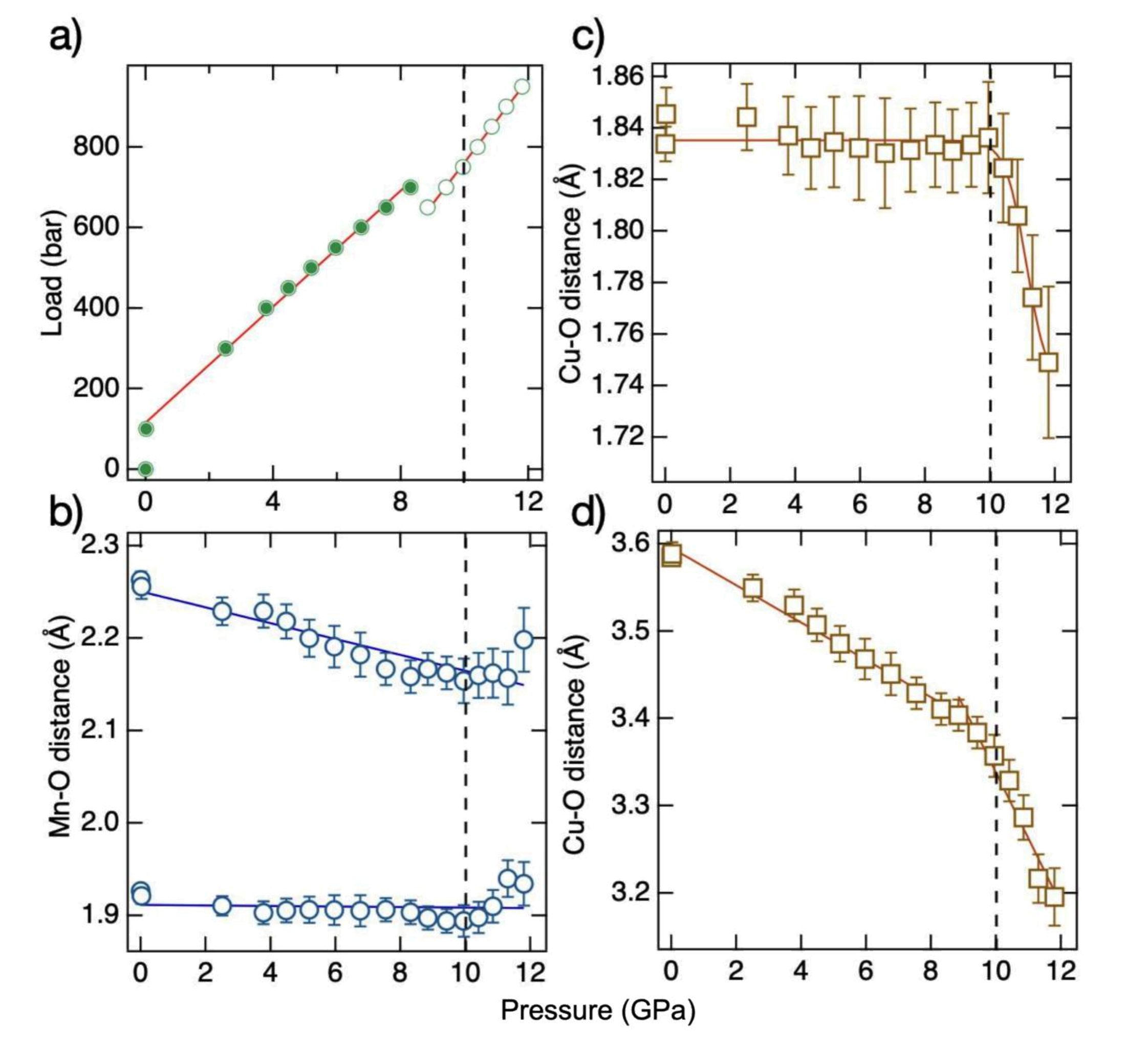}
	\caption{a) Load-pressure curves extracted for both loadings of the Paris-Edinburgh press using \cu~as the pressure calibrant. Note the anomaly between loadings occurs at a much lower pressure ($\sim$ 8 GPa) than the structural phase transition; b) Mn-O bond lengths as a function of pressure, note the undiminished Jahn-Teller distortion throughout; c) Pressure dependence of the Cu-O 'dumbell' bonds; d) Pressure dependence of the longer Cu-O contacts. Lines in the 0$<$P$<$8 GPa range represent fits used to extract compressibilities, and error bars are derived from Rietveld refinement against the experimental count-based errors. }
	\label{fgr:XRD}
\end{figure}

\subsection{Simulations}
The Jahn-Teller tetragonal elongation of the MnO$_6$ octahedra is present above the N\'{e}el temperature, 65\,K, showing the localized Mn d states are occupied in the paramagnetic state.
Previous measurements found no magneto-elastic distortions in the paramagnetic state,\cite{damay2009spin} so its phonons should be modellable with only the primitive unit cell. 
Paramagnetism with localized moments is difficult to simulate in general, impossible with such a cell. 
For computational ease, we employed a ferromagnetic (FM) representation as a proxy which maintains a high-spin (S=2) description of the Mn d electrons.\cite{Jia_MagFrust_2011}

\begin{figure}[]
	\includegraphics[width=7 cm]{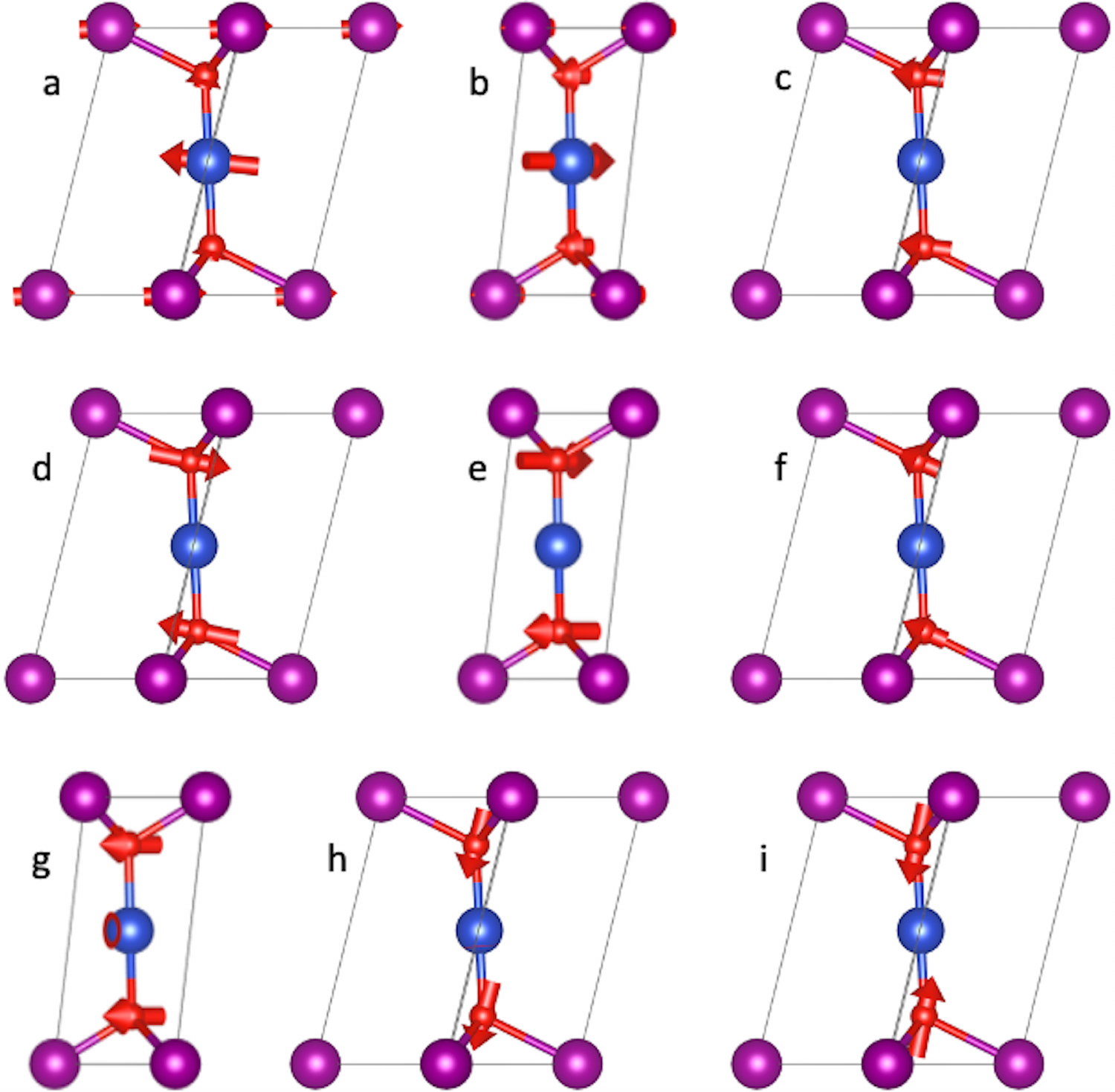}
	\caption{The DFT predicted optical phonon modes for a primitive unit cell of $C2/m$ CuMnO$_2$ in the experimental geometry at 0\,GPa as viewed along [110] (a,c,d,f,h,i) or [$\overline{1}$00] (b,e,g): a) 167 cm$^{-1}$ B$_{u}$, b) 207 cm$^{-1}$ A$_{u}$, c) 265 cm$^{-1}$ B$_{u}$, d) 329 cm$^{-1}$ A$_{g}$, e) 401 cm$^{-1}$ B$_{g}$, f) 423 cm$^{-1}$ B$_{u}$, g) 538 cm$^{-1}$ A$_{u}$, h) 692 cm$^{-1}$ B$_{u}$, i) 704 cm$^{-1}$ A$_{g}$. The Mn spins are ferromagnetically aligned to represent a single structural unit cell. The purple atoms are Mn, blue Cu, and red oxygen.}
	\label{fig:amb_mode}
\end{figure}

This model predicts the three $C2/m$~symmetry allowed\cite{Bilbao_SAM} and measured Raman active modes (Fig. 5) to have $\Gamma$-point frequencies of $A_g$: 329.1, $B_g$: 400.6, and $A_g$: 703.9\,cm$^{-1}$.
As illustrated in figure~\ref{fig:amb_mode}, the 703.9\,cm$^{-1}$ $A_g$ mode is an asymmetric stretch of the O--Cu--O unit.
The other modes are asymmetric displacements of the oxygen atoms along the primitive cell's [-110] and [110], for 329.1 and 400.6\,cm$^{-1}$ respectively.
In the full $C2/m$ cell these correspond to the [100] and [010] directions, respectively.
Relaxing the structure has a minimal impact on the lattice as the values of U$_\text{eff}$ were chosen to be similar to best reproduce the experimental lattice with this model: $a$=$b$ = 3.146 vs. 3.143\,\AA, $c$ = 5.872 vs. 5.884\,\AA, $\alpha$=$\beta$ = 77.712 vs. 77.305$^\circ$, and $\gamma$ = 54.447 vs. 55.087$^\circ$, for the experimental vs. DFT structure, respectively.
The relaxation does not qualitatively alter the atomic displacements along the modes, but it does shift the predicted Raman frequencies to $A_g$: 330.6, $B_g$: 369.1, and $A_g$: 694.9\,cm$^{-1}$.
While there is minimal change to the lattice upon optimization, the MnO$_6$ octahedra do become slightly more regularized with the difference in axial and equatorial bond length going from 0.337 to 0.300\,\AA.
This shows that the Raman active modes, in particular the $B_g$ mode, are sensitive to the positions of the oxygen atoms.

The lowest energy predicted optical mode is a 167.1\,cm$^{-1}$ (20.7 meV; 142.5\,cm$^{-1}$ in relaxed structure) $B_u$ mode which identifies the low energy feature seen in the inelastic neutron scattering (Fig. 4a) as a librational motion of the O--Cu--O unit.
Another librational $A_u$ mode is predicted at 207.0\,cm$^{-1}$ (204.8\,cm$^{-1}$ in relaxed structure), and its displacement is 90$^\circ$ to the 167.1\,cm$^{-1}$ mode.
These "guitar string" librational motions when coupled with increased bond lengths due to heating can lead to negative thermal expansion (NTE) via a covalency in either a rigid or molecular unit of the material.\cite{Miller2009,C8CC01153B}
These effects typically induce a bend in the rigid unit upon heating, but the librational mode can freely rotate in one plane, here about the Cu position in (002).\cite{Mast2019}
This makes the rigid unit still appear linear by diffraction but with large thermal ellipsoids, as is seen in figure~\ref{fgr:structure}.
Bonding analysis of the O--Cu--O$^{-3}$ unit shows a covalency between the copper and oxygen atoms, in agreement with estimates of charge transfer from NMR experiments\cite{zorko2015magnetic}. 
The individual polar $\sigma$ bonds between the copper and each oxygen have a 20\% occupation on the copper atom; this reduces to 10\% in O--Na--O$^{-3}$.
While most librations that lead to NTE involve a lighter atom tethered between two heavier atoms (ie. M--O--M), the O--Cu--O covalency should be sufficient to drive NTE via these librational modes.\cite{li2002strong,Lawler2017}

Compression beyond the phase transition to 11.8\,GPa (using the experimental structure) hardens the predicted $A_g$ Raman modes to 334.0 and 766.2\,cm$^{-1}$ and softens the $B_g$ mode to 287.9\,cm$^{-1}$.
The relaxed structure at 11.8\,GPa with this model is again similar to the experimental structure with the cell edges within 0.0056\,\AA{} and the cell angles within 0.63$^\circ$.
The key difference is the shape of the MnO$_6$ octahedra; the difference between the axial and equatorial bond lengths reduces from 0.263\,\AA{} in the experimental structure to 0.228\,\AA{} in the optimized structure.
Unlike at 0\,GPa, where this magnitude of change in the distortion of the MnO$_6$ octahedron softened the $B_g$ mode, the optimization hardens the $B_g$ mode to 351.5\,cm$^{-1}$.
The $A_g$ modes shift to 383.278 and 731.648 cm$^{-1}$.
Both Cu librational modes soften in the experimental (91.5 and 112.0\,cm$^{-1}$) and optimized (129.56 and 183.0\,cm$^{-1}$) 11.8\,GPa structures, signaling an increased propensity for NTE at higher pressures.

To investigate the magneto-elastic effects of CuMnO$_2$ in the antiferromagnetic (AFM) state, the previously discussed model must be expanded.
2$\times$1$\times$1 and 2$\times$1$\times$2 supercells of the primitive cell can model the magnetic propagation vectors of $\vec{\textbf{k}}=(-\frac{1}{2},\frac{1}{2},0)$ $\vec{\textbf{k}}=(-\frac{1}{2},\frac{1}{2},\frac{1}{2})$, respectively.\cite{damay2009spin,vecchini2010magnetoelastic,garlea2011tuning}
Both AFM representations produce very similar phonon frequencies and optimized lattices within 0.002\,\AA{} and 0.09$^{\circ}$, so we will only discuss the results of $\vec{\textbf{k}}=(-\frac{1}{2},\frac{1}{2},\frac{1}{2})$.
The Raman active phonon frequencies of the AFM representations of the experimental structures change by at most 5.1\,cm$^{-1}$.
However, the librational modes soften considerably to 83.8 and 115.7\,cm$^{-1}$ at 0\,GPa and 65.3 and 92.2\,cm$^{-1}$ at 11.8\,GPa, in better agreement with the low energy feature seen by neutron scattering.

More significant than the softening of the Cu librational modes, is that a structural optimization at both pressures breaks the $a=b$ symmetry of the $C2/m$ primitive cell, lowering the overall symmetry to $P\overline{1}$.
The structural distortions lowering the symmetry to $P\overline{1}$ can best be described as a contraction in the direction of nearest neighbor antiparallel spins, primarily [100] in the primitive cell ([110] in the full $C2/m$ structure) and to a lesser extent [001] (both cells).
This provides an explanation as to why the $a$ axis is seen to contract more than the $b$ axis upon initial compression (fig.~\ref{fgr:lattice-params}).
In the lower symmetry cell, the $A_g$, $B_g$, and $A_g$ Raman active modes (now all $A_g$ by symmetry) become 331.3, 377.6, and 693.8\,cm$^{-1}$ at 0\,GPa and 329.2, 406.7, and 729.2\,cm$^{-1}$ at 11.8\,GPa.
Following the conversion from $C2/m$ to $P\overline{1}$, the librational modes harden to 112.4 and 183.0\,cm$^{-1}$ at 0\,GPa and 107.3 and 194.0\,cm$^{-1}$ at 11.8\,GPa. This shows that the phase transition is driven by the softening of the librational modes which is precipitated primarily by antiferromagnetic ordering and secondarily pressure.

\section{Discussion}
The above results show that the dynamics of the linear O-Cu-O units strongly influence the structure of \cu. This is equally true at ambient and high pressures. In the former case, as has been reported for many other delafossite materials\cite{li2002strong,li2005trends,ahmed2009negative}, transverse 'guitar string' modes are populated upon heating. These draw together the MnO$_{6}$~planes, and are manifested in the aniostropic thermal factors and low energy 11.9 meV phonon excitation. Assignment of this mode to transverse Cu vibrations can be made with  confidence by comparison with other delafossites\cite{abramchuk2018crystal}~containing interlayer Cu, and the agreement with the Einstein oscillator fit to the Cu~$U_{11}$~displacements parameters, as well as our DFT simulations. In addition, the  observed phonon intensity is dominated by copper due to its large scattering cross section. In a  simple picture, the $c$-axis NTE should cease when this mode is completely thermally populated, assuming no other interactions come into play. Experimentally, this occurs at $\sim$180 K, corresponding to a Boltzmann factor $p_{i}/p_{j}=exp(-\Delta E/k_{B}T)$=0.46 (Fig. 1b). This picture is thus consistent with the ambient pressure thermal expansion to zeroth order. That said, there are profound differences with non-magnetic delafossites. From the available literature\cite{li2002strong,li2005trends,ahmed2009negative}, we were unable to find any examples of Cu delafossites with residual disorder of the anisotropic ADPs. Furthermore, the almost 1D nature of the displacements (which are perpendicular to the strongest magnetic coupling) have no counterpart which we are aware of. The nearly temperature independent values of the ADPs below $T_{N}$=65 K, offer a possible explanation. Upon cooling \na, magnetic frustration drives local symmetry breaking, which manifests as short-range order in neutron scattering\cite{frandsen2020nanoscale}, as well as increased lattice microstrain\cite{zorko2014frustration}. However, neither our X-ray PDF measurements (Fig. 3), nor reported neutron PDFs\cite{frandsen2020nanoscale}~show any such effect for \cu. These experiments used structure models in $C2/m$, with all symmetry allowed variables refined. This includes the Cu ADPs. It was hence concluded that stoichiometric \cu~is below a critical disorder limit for the anisotropic triangular lattice. Our interpretation is that magnetic frustration is indeed coupled to short range lattice displacements. However, these can be parameterized by a symmetry allowed crystallographic degree of freedom.  This might also explain the broad linewidths observed by $^{63,65}$Cu NMR and NQR spectra\cite{zorko2015magnetic}. The energy match between the low-Q magnetism and the flat phonon mode (Fig. 4) allows hybridisation, and hence enhanced spin-lattice coupling through this channel.\\
The development of the structure of \cu~at P $<$10 GPa is somewhat similar to other Cu delafossites, although the bulk modulus of 116(2) GPa makes it the softest material in this class\cite{garg2018copper}. In all of these materials, the $c$-axis is initially the stiffest, attesting to the covalent nature of the O-Cu-O bonding. Indeed, the $A_{g}$~vibration is also ubiquitously found to harden\cite{Ramancuga,Ramancual,garg2014multiferroic,salke2015raman}~on compression. Above 10 GPa, the response of \cu~diverges from non-magnetic Cu delafossites. The changes, which include the pronounced collapse of the $c$-axis and the switch in compressibility between the $a$~and $b$~axes, have many possible explanations. Using only X-ray diffraction, we are unable to discriminate between coordination changes, charge-transfer, orbital effects etc. This arises due to the intrinsic insensitivity of this technique to oxide positions, as well as problems with the microstructure of our sample. As we discuss below, the lattice changes can in fact also be linked to the same O-Cu-O vibrational degrees of freedom which template the ambient pressure behaviour. This requires an understanding of the limits to Rietveld analysis, and complimentary use of the Raman and simulation results. We begin our discussions by reviewing the relevant experimental observations.\\
The most important observation from our X-ray diffraction experiments at high pressure is the isostructural nature of the 10 GPa phase transition. We detected no evidence for lattice distortions, nor any change in the number of peaks. This is completely consistent with the unchanged number of modes in the Raman spectra across the transition. We also note that these experiments showed no decrease in signal, which could be indicative of increased reflectivity and metallisation. This is quite unlike the charge-transfer transition in CuFeO$_{2}$, which leads to a change in lattice symmetry from monocolinic to trigonal in a fraction of the sample\cite{xu2010pressure}, as well as a wipe-out of the Raman modes\cite{salke2015raman}. \\
At first sight the neutron powder diffraction results shown in Fig. 9 conflict  with this conclusion, as the reduction in Cu-O bond lengths and uptick in Mn-O bonds lengths is exactly what would be expected from charge transfer. We believe that we can reconcile the facts as follows: 1) If the Cu-O bond lengths were genuinely undergoing a gross reduction, and especially if the Cu was oxidised, the O-Cu-O $A_{g}$~stretch measured by Raman would harden, not soften at 10 GPa (Fig. 5, bottom); 2) The bond distance measured by diffraction depends upon only the time averaged positions of the atoms involved. If the instantaneous positions of the atoms never pass through this point, erroneous results may be found\cite{busing1964effect}. This is especially common for classic NTE materials with pronounced librational motions\cite{chapman2005direct, chapman2009anomalous}. While we do not directly probe the instantaneous correlations in this work, our DFT simulations strongly support our conclusions. Relaxing the $C2/m$~structure at ambient or high pressure gives the same result- bent O-Cu-O linkages. The resulting Cu-O bond length  is only slightly reduced at high pressure, consistent with the overall reduction in cell volume. The 10 GPa isostructural transition in \cu~is thus driven by an increase of transverse Cu motion, and the other effects (change in cell axis compressibilities etc) are secondary. Finally, we note that a very recent work \cite{levy2020high} has examined CuMnO$_{2}$ under pressure using X-ray absorption spectroscopy. These results also rule out charge transfer, consistent with our results.\\
Before concluding this work, we note that there are some questions which remain unanswered. For example, it is interesting that this transition occurs abruptly, with a pronounced volume anomaly (Fig. 3b). In the cartoon model developed above, there is no reason for the $c$-axis not to contract smoothly under pressure. Similar to the minimum in this cell axis seen at ambient pressure around 180 K, it is likely that this reflects other longer range interactions. Future studies will be needed to determine if these are material specific, or if they have more general implications for a broader class of materials. Finally, the observation of local disorder raises interesting questions about the magnetic frustration in \na~and \cu. Both materials are now shown to have local atomic displacements, which contribute to breaking magnetic frustration. Similar effects have been reported in frustrated pyrochlores\cite{keren2001frustration}, dimer systems\cite{kimber2014valence}~and charge-ordering materials\cite{perversi2019co}. Future crystallographic or diffuse scattering experiments on \cu~will be needed to follow these as a function of temperature, and perhaps to determine any scaling with the magnetic order parameter.\\

\section{Conclusions}
In summary, we have reported a thorough characterisation of the structure and dynamics of the delafossite \cu~as a function of temperature and pressure. We report the discovery of positional disorder of the interlayer Cu$^{+}$~cations, as parameterised by a non-zero ADP value. This is found to be temperature independent below the magnetic ordering temperature. The frustration breaking structural phase transition and the $c$-axis negative thermal expansion are thus both related to anisotropic motion. This is reflected in the low-lying 11.9 meV phonon mode, and our DFT simulations. Under high-pressures, we have discovered an isostructural phase transition at 10 GPa, where the $c$-axis collapses. Base on our results and simulations, this appears to be driven by the same dynamical degree of freedom.\\

\section{Acknowledgements}
The high-pressure part of this project originated during the post-doctoral employment of S.R.E, A.S and S.A.J.K at the European Synchrotron Radiation Facility, which we acknowledge for access to beam time. In addition, we thank Michael Hanfland for loading the pressure cell used on ID09A, Pam Whitfield for help with \textsc{TOPAS}, and Clemens Ritter for assistance on D20. This research used resources at the Spallation Neutron Source, a DOE Office of Science User Facility operated by the Oak Ridge National Laboratory. Oak Ridge National Laboratory is managed by UT-Batelle, LLC, for the DOE under contract DE-AC05-1008 00OR22725. This research was sponsored in part by the National Nuclear Security Administration under the Stewardship Science Academic Alliances program through DOE Co-operative Agreement DE-NA0001982. We thank the anonymous referees for their many useful comments. Ce travail a \'{e}t\'{e} soutenu par le programme "Investissements d'Avenir", projet ISITE-BFC (contrat ANR-15-IDEX-0003).





\bibliographystyle{acm}
\bibliography{CuMnO2_TC}

\providecommand{\latin}[1]{#1}
\makeatletter
\providecommand{\doi}
  {\begingroup\let\do\@makeother\dospecials
  \catcode`\{=1 \catcode`\}=2 \doi@aux}
\providecommand{\doi@aux}[1]{\endgroup\texttt{#1}}
\makeatother
\providecommand*\mcitethebibliography{\thebibliography}
\csname @ifundefined\endcsname{endmcitethebibliography}
  {\let\endmcitethebibliography\endthebibliography}{}
\begin{mcitethebibliography}{75}
\providecommand*\natexlab[1]{#1}
\providecommand*\mciteSetBstSublistMode[1]{}
\providecommand*\mciteSetBstMaxWidthForm[2]{}
\providecommand*\mciteBstWouldAddEndPuncttrue
  {\def\EndOfBibitem{\unskip.}}
\providecommand*\mciteBstWouldAddEndPunctfalse
  {\let\EndOfBibitem\relax}
\providecommand*\mciteSetBstMidEndSepPunct[3]{}
\providecommand*\mciteSetBstSublistLabelBeginEnd[3]{}
\providecommand*\EndOfBibitem{}
\mciteSetBstSublistMode{f}
\mciteSetBstMaxWidthForm{subitem}{(\alph{mcitesubitemcount})}
\mciteSetBstSublistLabelBeginEnd
  {\mcitemaxwidthsubitemform\space}
  {\relax}
  {\relax}

\bibitem[Marquardt \latin{et~al.}(2006)Marquardt, Ashmore, and
  Cann]{marquardt2006crystal}
Marquardt,~M.~A.; Ashmore,~N.~A.; Cann,~D.~P. Crystal chemistry and electrical
  properties of the delafossite structure. \emph{Thin Solid Films}
  \textbf{2006}, \emph{496}, 146--156\relax
\mciteBstWouldAddEndPuncttrue
\mciteSetBstMidEndSepPunct{\mcitedefaultmidpunct}
{\mcitedefaultendpunct}{\mcitedefaultseppunct}\relax
\EndOfBibitem
\bibitem[Goodenough \latin{et~al.}(1991)Goodenough, Dutta, and
  Manthiram]{goodenough1991lattice}
Goodenough,~J.; Dutta,~G.; Manthiram,~A. Lattice instabilities near the
  critical V-V separation for localized versus itinerant electrons in LiV$_{1-
  y}$M$_{y}$O$_{2}$ (M= Cr or Ti) Li$_{1-x}$VO$_2$. \emph{Phys. Rev. B}
  \textbf{1991}, \emph{43}, 10170\relax
\mciteBstWouldAddEndPuncttrue
\mciteSetBstMidEndSepPunct{\mcitedefaultmidpunct}
{\mcitedefaultendpunct}{\mcitedefaultseppunct}\relax
\EndOfBibitem
\bibitem[Onoda and Inabe(1993)Onoda, and Inabe]{onoda1993role}
Onoda,~M.; Inabe,~T. Role of Structural Change in Phase Transition in LiVO$_2$.
  \emph{J. Phys. Soc. Japan} \textbf{1993}, \emph{62}, 2216--2219\relax
\mciteBstWouldAddEndPuncttrue
\mciteSetBstMidEndSepPunct{\mcitedefaultmidpunct}
{\mcitedefaultendpunct}{\mcitedefaultseppunct}\relax
\EndOfBibitem
\bibitem[Frontzek \latin{et~al.}(2011)Frontzek, Ehlers, Podlesnyak, Cao,
  Matsuda, Zaharko, Aliouane, Barilo, and Shiryaev]{frontzek2011}
Frontzek,~M.; Ehlers,~G.; Podlesnyak,~A.; Cao,~H.; Matsuda,~M.; Zaharko,~O.;
  Aliouane,~N.; Barilo,~S.; Shiryaev,~S. Magnetic structure of CuCrO$_2$: a
  single crystal neutron diffraction study. \emph{J. Phys. Cond. Matt.}
  \textbf{2011}, \emph{24}, 016004\relax
\mciteBstWouldAddEndPuncttrue
\mciteSetBstMidEndSepPunct{\mcitedefaultmidpunct}
{\mcitedefaultendpunct}{\mcitedefaultseppunct}\relax
\EndOfBibitem
\bibitem[Stock \latin{et~al.}(2009)Stock, Chapon, Adamopoulos, Lappas, Giot,
  Taylor, Green, Brown, and Radaelli]{stock2009one}
Stock,~C.; Chapon,~L.; Adamopoulos,~O.; Lappas,~A.; Giot,~M.; Taylor,~J.;
  Green,~M.; Brown,~C.; Radaelli,~P. One-Dimensional Magnetic Fluctuations in
  the Spin-2 Triangular Lattice $\alpha$- NaMnO$_2$. \emph{Phys. Rev. Lett.}
  \textbf{2009}, \emph{103}, 077202\relax
\mciteBstWouldAddEndPuncttrue
\mciteSetBstMidEndSepPunct{\mcitedefaultmidpunct}
{\mcitedefaultendpunct}{\mcitedefaultseppunct}\relax
\EndOfBibitem
\bibitem[Dally \latin{et~al.}(2018)Dally, Zhao, Xu, Chisnell, Stone, Lynn,
  Balents, and Wilson]{dally2018amplitude}
Dally,~R.~L.; Zhao,~Y.; Xu,~Z.; Chisnell,~R.; Stone,~M.~B.; Lynn,~J.~W.;
  Balents,~L.; Wilson,~S.~D. Amplitude mode in the planar triangular
  antiferromagnet Na$_{0.9}$MnO$_2$. \emph{Nat. Commun.} \textbf{2018},
  \emph{9}, 1--8\relax
\mciteBstWouldAddEndPuncttrue
\mciteSetBstMidEndSepPunct{\mcitedefaultmidpunct}
{\mcitedefaultendpunct}{\mcitedefaultseppunct}\relax
\EndOfBibitem
\bibitem[Kimber \latin{et~al.}(2020)Kimber, Wildes, Mutka, Bos, and
  Argyriou]{kimber2020spin}
Kimber,~S.~A.; Wildes,~A.~R.; Mutka,~H.; Bos,~J.-W.~G.; Argyriou,~D.~N.
  Spin-chain correlations in the frustrated triangular lattice material
  CuMnO$_2$. \emph{J. Phys. Cond. Matt.} \textbf{2020}, \emph{32}, 445802\relax
\mciteBstWouldAddEndPuncttrue
\mciteSetBstMidEndSepPunct{\mcitedefaultmidpunct}
{\mcitedefaultendpunct}{\mcitedefaultseppunct}\relax
\EndOfBibitem
\bibitem[Giot \latin{et~al.}(2007)Giot, Chapon, Androulakis, Green, Radaelli,
  and Lappas]{giot2007magnetoelastic}
Giot,~M.; Chapon,~L.~C.; Androulakis,~J.; Green,~M.~A.; Radaelli,~P.~G.;
  Lappas,~A. Magnetoelastic Coupling and Symmetry Breaking in the Frustrated
  Antiferromagnet $\alpha$- NaMnO$_2$. \emph{Phys. Rev. Lett.} \textbf{2007},
  \emph{99}, 247211\relax
\mciteBstWouldAddEndPuncttrue
\mciteSetBstMidEndSepPunct{\mcitedefaultmidpunct}
{\mcitedefaultendpunct}{\mcitedefaultseppunct}\relax
\EndOfBibitem
\bibitem[Zorko \latin{et~al.}(2014)Zorko, Adamopoulos, Komelj, Ar{\v{c}}on, and
  Lappas]{zorko2014frustration}
Zorko,~A.; Adamopoulos,~O.; Komelj,~M.; Ar{\v{c}}on,~D.; Lappas,~A.
  Frustration-induced nanometre-scale inhomogeneity in a triangular
  antiferromagnet. \emph{Nat. Commun.} \textbf{2014}, \emph{5}, 3222\relax
\mciteBstWouldAddEndPuncttrue
\mciteSetBstMidEndSepPunct{\mcitedefaultmidpunct}
{\mcitedefaultendpunct}{\mcitedefaultseppunct}\relax
\EndOfBibitem
\bibitem[Damay \latin{et~al.}(2009)Damay, Poienar, Martin, Maignan,
  Rodriguez-Carvajal, Andr\'e, and Doumerc]{damay2009spin}
Damay,~F.; Poienar,~M.; Martin,~C.; Maignan,~A.; Rodriguez-Carvajal,~J.;
  Andr\'e,~G.; Doumerc,~J.~P. {Spin-Lattice Coupling Induced Phase Transition
  in the $S=2$ Frustrated Antiferromagnet ${\text{CuMnO}}_{2}$}. \emph{Phys.
  Rev. B} \textbf{2009}, \emph{80}, 094410, DOI:
  \doi{10.1103/PhysRevB.80.094410}\relax
\mciteBstWouldAddEndPuncttrue
\mciteSetBstMidEndSepPunct{\mcitedefaultmidpunct}
{\mcitedefaultendpunct}{\mcitedefaultseppunct}\relax
\EndOfBibitem
\bibitem[Vecchini \latin{et~al.}(2010)Vecchini, Poienar, Damay, Adamopoulos,
  Daoud-Aladine, Lappas, Perez-Mato, Chapon, and
  Martin]{vecchini2010magnetoelastic}
Vecchini,~C.; Poienar,~M.; Damay,~F.; Adamopoulos,~O.; Daoud-Aladine,~A.;
  Lappas,~A.; Perez-Mato,~J.~M.; Chapon,~L.~C.; Martin,~C. {Magnetoelastic
  Coupling in the Frustrated Antiferromagnetic Triangular Lattice
  ${\text{CuMnO}}_{2}$}. \emph{Phys. Rev. B} \textbf{2010}, \emph{82}, 094404,
  DOI: \doi{10.1103/PhysRevB.82.094404}\relax
\mciteBstWouldAddEndPuncttrue
\mciteSetBstMidEndSepPunct{\mcitedefaultmidpunct}
{\mcitedefaultendpunct}{\mcitedefaultseppunct}\relax
\EndOfBibitem
\bibitem[Terada \latin{et~al.}(2011)Terada, Tsuchiya, Kitazawa, Osakabe,
  Metoki, Igawa, and Ohoyama]{terada2011magnetic}
Terada,~N.; Tsuchiya,~Y.; Kitazawa,~H.; Osakabe,~T.; Metoki,~N.; Igawa,~N.;
  Ohoyama,~K. Magnetic correlations and the influence of atomic disorder in
  frustrated isosceles triangular lattice antiferromagnet CuMnO$_2$.
  \emph{Phys. Rev. B} \textbf{2011}, \emph{84}, 064432\relax
\mciteBstWouldAddEndPuncttrue
\mciteSetBstMidEndSepPunct{\mcitedefaultmidpunct}
{\mcitedefaultendpunct}{\mcitedefaultseppunct}\relax
\EndOfBibitem
\bibitem[Frandsen \latin{et~al.}(2020)Frandsen, Bozin, Aza, Mart{\'\i}nez,
  Feygenson, Page, and Lappas]{frandsen2020nanoscale}
Frandsen,~B.~A.; Bozin,~E.~S.; Aza,~E.; Mart{\'\i}nez,~A.~F.; Feygenson,~M.;
  Page,~K.; Lappas,~A. Nanoscale degeneracy lifting in a geometrically
  frustrated antiferromagnet. \emph{Phys. Rev. B} \textbf{2020}, \emph{101},
  024423\relax
\mciteBstWouldAddEndPuncttrue
\mciteSetBstMidEndSepPunct{\mcitedefaultmidpunct}
{\mcitedefaultendpunct}{\mcitedefaultseppunct}\relax
\EndOfBibitem
\bibitem[Garlea \latin{et~al.}(2011)Garlea, Savici, and Jin]{garlea2011tuning}
Garlea,~V.~O.; Savici,~A.~T.; Jin,~R. Tuning the magnetic ground state of a
  triangular lattice system Cu(Mn$_{1-x}$Cu$_x$)O$_2$. \emph{Phys. Rev. B}
  \textbf{2011}, \emph{83}, 172407\relax
\mciteBstWouldAddEndPuncttrue
\mciteSetBstMidEndSepPunct{\mcitedefaultmidpunct}
{\mcitedefaultendpunct}{\mcitedefaultseppunct}\relax
\EndOfBibitem
\bibitem[Poienar \latin{et~al.}(2011)Poienar, Vecchini, Andre, Daoud-Aladine,
  Margiolaki, Maignan, Lappas, Chapon, Hervieu, and
  Damay]{poienar2011substitution}
Poienar,~M.; Vecchini,~C.; Andre,~G.; Daoud-Aladine,~A.; Margiolaki,~I.;
  Maignan,~A.; Lappas,~A.; Chapon,~L.; Hervieu,~M.; Damay,~F. Substitution
  effect on the interplane coupling in crednerite: the Cu$_{1.04}$Mn$_{0.96O2}$
  case. \emph{Chem. Mater,} \textbf{2011}, \emph{23}, 85--94\relax
\mciteBstWouldAddEndPuncttrue
\mciteSetBstMidEndSepPunct{\mcitedefaultmidpunct}
{\mcitedefaultendpunct}{\mcitedefaultseppunct}\relax
\EndOfBibitem
\bibitem[Zorko \latin{et~al.}(2015)Zorko, Kokalj, Komelj, Adamopoulos,
  Luetkens, Ar{\v{c}}on, and Lappas]{zorko2015magnetic}
Zorko,~A.; Kokalj,~J.; Komelj,~M.; Adamopoulos,~O.; Luetkens,~H.;
  Ar{\v{c}}on,~D.; Lappas,~A. Magnetic inhomogeneity on a triangular lattice:
  the magnetic-exchange versus the elastic energy and the role of disorder.
  \emph{Scientific reports} \textbf{2015}, \emph{5}, 1--8\relax
\mciteBstWouldAddEndPuncttrue
\mciteSetBstMidEndSepPunct{\mcitedefaultmidpunct}
{\mcitedefaultendpunct}{\mcitedefaultseppunct}\relax
\EndOfBibitem
\bibitem[Li \latin{et~al.}(2002)Li, Yokochi, Amos, and Sleight]{li2002strong}
Li,~J.; Yokochi,~A.; Amos,~T.; Sleight,~A. Strong Negative Thermal Expansion
  along the O-Cu-O Linkage in CuScO$_2$. \emph{Chem. Mater,} \textbf{2002},
  \emph{14}, 2602--2606\relax
\mciteBstWouldAddEndPuncttrue
\mciteSetBstMidEndSepPunct{\mcitedefaultmidpunct}
{\mcitedefaultendpunct}{\mcitedefaultseppunct}\relax
\EndOfBibitem
\bibitem[Li \latin{et~al.}(2005)Li, Sleight, Jones, and Toby]{li2005trends}
Li,~J.; Sleight,~A.; Jones,~C.; Toby,~B. Trends in negative thermal expansion
  behavior for AMO$_2$ (A= Cu or Ag; M= Al, Sc, In, or La) compounds with the
  delafossite structure. \emph{J. Sol. State Chem.} \textbf{2005}, \emph{178},
  285--294\relax
\mciteBstWouldAddEndPuncttrue
\mciteSetBstMidEndSepPunct{\mcitedefaultmidpunct}
{\mcitedefaultendpunct}{\mcitedefaultseppunct}\relax
\EndOfBibitem
\bibitem[Ahmed \latin{et~al.}(2009)Ahmed, Dalba, Fornasini, Vaccari, Rocca,
  Sanson, Li, and Sleight]{ahmed2009negative}
Ahmed,~S.; Dalba,~G.; Fornasini,~P.; Vaccari,~M.; Rocca,~F.; Sanson,~A.;
  Li,~J.; Sleight,~A. Negative thermal expansion in crystals with the
  delafossite structure: An extended X-ray absorption fine structure study of
  CuScO$_2$ and CuLaO$_2$. \emph{Phys. Rev. B} \textbf{2009}, \emph{79},
  104302\relax
\mciteBstWouldAddEndPuncttrue
\mciteSetBstMidEndSepPunct{\mcitedefaultmidpunct}
{\mcitedefaultendpunct}{\mcitedefaultseppunct}\relax
\EndOfBibitem
\bibitem[Zhao \latin{et~al.}(1997)Zhao, Hasegawa, Kondo, Yagi, and
  Takei]{zhao1997x}
Zhao,~T.; Hasegawa,~M.; Kondo,~T.; Yagi,~T.; Takei,~H. X-ray diffraction study
  of copper iron oxide [CuFeO$_2$] under pressures up to 10 GPa. \emph{Mat.
  Res. Bull.} \textbf{1997}, \emph{32}, 151--157\relax
\mciteBstWouldAddEndPuncttrue
\mciteSetBstMidEndSepPunct{\mcitedefaultmidpunct}
{\mcitedefaultendpunct}{\mcitedefaultseppunct}\relax
\EndOfBibitem
\bibitem[Xu \latin{et~al.}(2010)Xu, Rozenberg, Pasternak, Kertzer, Kurnosov,
  Dubrovinsky, Pascarelli, Munoz, Vaccari, Hanfland, and
  Jeanloz]{xu2010pressure}
Xu,~W.~M.; Rozenberg,~G.~K.; Pasternak,~M.~P.; Kertzer,~M.; Kurnosov,~A.;
  Dubrovinsky,~L.~S.; Pascarelli,~S.; Munoz,~M.; Vaccari,~M.; Hanfland,~M.;
  Jeanloz,~R. {Pressure-Induced
  $\text{Fe}\ensuremath{\leftrightarrow}\text{Cu}$ Cationic Valence Exchange
  and Its Structural Consequences: High-Pressure Studies of Delafossite
  ${\text{CuFeO}}_{2}$}. \emph{Phys. Rev. B} \textbf{2010}, \emph{81}, 104110,
  DOI: \doi{10.1103/PhysRevB.81.104110}\relax
\mciteBstWouldAddEndPuncttrue
\mciteSetBstMidEndSepPunct{\mcitedefaultmidpunct}
{\mcitedefaultendpunct}{\mcitedefaultseppunct}\relax
\EndOfBibitem
\bibitem[Salke \latin{et~al.}(2015)Salke, Kamali, Ravindran, Balakrishnan, and
  Rao]{salke2015raman}
Salke,~N.~P.; Kamali,~K.; Ravindran,~T.; Balakrishnan,~G.; Rao,~R. Raman
  spectroscopic studies of CuFeO$_2$ at high pressures. \emph{Vibrational
  Spectroscopy} \textbf{2015}, \emph{81}, 112--118\relax
\mciteBstWouldAddEndPuncttrue
\mciteSetBstMidEndSepPunct{\mcitedefaultmidpunct}
{\mcitedefaultendpunct}{\mcitedefaultseppunct}\relax
\EndOfBibitem
\bibitem[Xu \latin{et~al.}(2016)Xu, Hearne, and Pasternak]{xu2016cufe}
Xu,~W.; Hearne,~G.; Pasternak,~M. CuFeO$_2$ at a megabar: Stabilization of a
  mixed-valence low-spin magnetic semiconducting ground state. \emph{Phys. Rev.
  B} \textbf{2016}, \emph{94}, 035155\relax
\mciteBstWouldAddEndPuncttrue
\mciteSetBstMidEndSepPunct{\mcitedefaultmidpunct}
{\mcitedefaultendpunct}{\mcitedefaultseppunct}\relax
\EndOfBibitem
\bibitem[Garg and Rao(2018)Garg, and Rao]{garg2018copper}
Garg,~A.~B.; Rao,~R. Copper delafossites under high pressure- A brief review of
  XRD and Raman spectroscopic studies. \emph{Crystals} \textbf{2018}, \emph{8},
  255\relax
\mciteBstWouldAddEndPuncttrue
\mciteSetBstMidEndSepPunct{\mcitedefaultmidpunct}
{\mcitedefaultendpunct}{\mcitedefaultseppunct}\relax
\EndOfBibitem
\bibitem[T{\"o}bbens \latin{et~al.}(2001)T{\"o}bbens, St{\"u}{\ss}er, Knorr,
  Mayer, and Lampert]{tobbens2001e9}
T{\"o}bbens,~D.; St{\"u}{\ss}er,~N.; Knorr,~K.; Mayer,~H.; Lampert,~G. E9: the
  new high-resolution neutron powder diffractometer at the Berlin neutron
  scattering center. Materials Science Forum. 2001; pp 288--293\relax
\mciteBstWouldAddEndPuncttrue
\mciteSetBstMidEndSepPunct{\mcitedefaultmidpunct}
{\mcitedefaultendpunct}{\mcitedefaultseppunct}\relax
\EndOfBibitem
\bibitem[Toby and Von~Dreele(2013)Toby, and Von~Dreele]{toby2013gsas}
Toby,~B.~H.; Von~Dreele,~R.~B. GSAS-II: the genesis of a modern open-source all
  purpose crystallography software package. \emph{J. Appl. Cryst.}
  \textbf{2013}, \emph{46}, 544--549\relax
\mciteBstWouldAddEndPuncttrue
\mciteSetBstMidEndSepPunct{\mcitedefaultmidpunct}
{\mcitedefaultendpunct}{\mcitedefaultseppunct}\relax
\EndOfBibitem
\bibitem[Ashiotis \latin{et~al.}(2015)Ashiotis, Deschildre, Nawaz, Wright,
  Karkoulis, Picca, and Kieffer]{ashiotis2015fast}
Ashiotis,~G.; Deschildre,~A.; Nawaz,~Z.; Wright,~J.~P.; Karkoulis,~D.;
  Picca,~F.~E.; Kieffer,~J. The fast azimuthal integration Python library:
  pyFAI. \emph{J. Appl. Cryst.} \textbf{2015}, \emph{48}, 510--519\relax
\mciteBstWouldAddEndPuncttrue
\mciteSetBstMidEndSepPunct{\mcitedefaultmidpunct}
{\mcitedefaultendpunct}{\mcitedefaultseppunct}\relax
\EndOfBibitem
\bibitem[Juh{\'a}s \latin{et~al.}(2013)Juh{\'a}s, Davis, Farrow, and
  Billinge]{juhas2013pdfgetx3}
Juh{\'a}s,~P.; Davis,~T.; Farrow,~C.~L.; Billinge,~S.~J. PDFgetX3: a rapid and
  highly automatable program for processing powder diffraction data into total
  scattering pair distribution functions. \emph{J. Appl. Cryst.} \textbf{2013},
  \emph{46}, 560--566\relax
\mciteBstWouldAddEndPuncttrue
\mciteSetBstMidEndSepPunct{\mcitedefaultmidpunct}
{\mcitedefaultendpunct}{\mcitedefaultseppunct}\relax
\EndOfBibitem
\bibitem[Farrow \latin{et~al.}(2007)Farrow, Juhas, Liu, Bryndin, Bo{\v{z}}in,
  Bloch, Proffen, and Billinge]{farrow2007pdffit2}
Farrow,~C.; Juhas,~P.; Liu,~J.; Bryndin,~D.; Bo{\v{z}}in,~E.; Bloch,~J.;
  Proffen,~T.; Billinge,~S. PDFfit2 and PDFgui: computer programs for studying
  nanostructure in crystals. \emph{J. Phys. Cond. Matt.} \textbf{2007},
  \emph{19}, 335219\relax
\mciteBstWouldAddEndPuncttrue
\mciteSetBstMidEndSepPunct{\mcitedefaultmidpunct}
{\mcitedefaultendpunct}{\mcitedefaultseppunct}\relax
\EndOfBibitem
\bibitem[F{\aa}k and Dorner(1997)F{\aa}k, and Dorner]{faak1997phonon}
F{\aa}k,~B.; Dorner,~B. Phonon line shapes and excitation energies.
  \emph{Physica B} \textbf{1997}, \emph{234}, 1107--1108\relax
\mciteBstWouldAddEndPuncttrue
\mciteSetBstMidEndSepPunct{\mcitedefaultmidpunct}
{\mcitedefaultendpunct}{\mcitedefaultseppunct}\relax
\EndOfBibitem
\bibitem[Krimmel \latin{et~al.}(2009)Krimmel, Mutka, Koza, Tsurkan, and
  Loidl]{Loidl}
Krimmel,~A.; Mutka,~H.; Koza,~M.; Tsurkan,~V.; Loidl,~A. Spin excitations in
  frustrated A-site spinels investigated with inelastic neutron scattering.
  \emph{Phys. Rev. B} \textbf{2009}, \emph{79}, 134406\relax
\mciteBstWouldAddEndPuncttrue
\mciteSetBstMidEndSepPunct{\mcitedefaultmidpunct}
{\mcitedefaultendpunct}{\mcitedefaultseppunct}\relax
\EndOfBibitem
\bibitem[Hammersley(2016)]{hammersley2016fit2d}
Hammersley,~A. FIT2D: a multi-purpose data reduction, analysis and
  visualization program. \emph{J. Appl. Cryst.} \textbf{2016}, \emph{49},
  646--652\relax
\mciteBstWouldAddEndPuncttrue
\mciteSetBstMidEndSepPunct{\mcitedefaultmidpunct}
{\mcitedefaultendpunct}{\mcitedefaultseppunct}\relax
\EndOfBibitem
\bibitem[Larson and Von~Dreele(1994)Larson, and Von~Dreele]{larson1994gsas}
Larson,~A.~C.; Von~Dreele,~R.~B. GSAS. \emph{Report lAUR} \textbf{1994},
  86--748\relax
\mciteBstWouldAddEndPuncttrue
\mciteSetBstMidEndSepPunct{\mcitedefaultmidpunct}
{\mcitedefaultendpunct}{\mcitedefaultseppunct}\relax
\EndOfBibitem
\bibitem[Toby(2001)]{toby2001expgui}
Toby,~B.~H. EXPGUI, a graphical user interface for GSAS. \emph{J. Appl. Cryst.}
  \textbf{2001}, \emph{34}, 210--213\relax
\mciteBstWouldAddEndPuncttrue
\mciteSetBstMidEndSepPunct{\mcitedefaultmidpunct}
{\mcitedefaultendpunct}{\mcitedefaultseppunct}\relax
\EndOfBibitem
\bibitem[Calder \latin{et~al.}(2018)Calder, An, Boehler, Dela~Cruz, Frontzek,
  Guthrie, Haberl, Huq, Kimber, Liu, Molaison, Neuefeind, Page, dos Santos,
  Taddei, Tulk, and Tucker]{calder2018suite}
Calder,~S. \latin{et~al.}  {A Suite-Level Review of the Neutron Powder
  Diffraction Instruments at Oak Ridge National Laboratory}. \emph{Rev. Sci.
  Inst.} \textbf{2018}, \emph{89}, 092701, DOI: \doi{10.1063/1.5033906}\relax
\mciteBstWouldAddEndPuncttrue
\mciteSetBstMidEndSepPunct{\mcitedefaultmidpunct}
{\mcitedefaultendpunct}{\mcitedefaultseppunct}\relax
\EndOfBibitem
\bibitem[Besson and Nelmes(1995)Besson, and Nelmes]{besson1995new}
Besson,~J.; Nelmes,~R. New developments in neutron-scattering methods under
  high pressure with the Paris—Edinburgh cells. \emph{Physica B}
  \textbf{1995}, \emph{213}, 31--36\relax
\mciteBstWouldAddEndPuncttrue
\mciteSetBstMidEndSepPunct{\mcitedefaultmidpunct}
{\mcitedefaultendpunct}{\mcitedefaultseppunct}\relax
\EndOfBibitem
\bibitem[Khvostantsev \latin{et~al.}(2004)Khvostantsev, Slesarev, and
  Brazhkin]{Khvostantsev2004_DT_Anvils}
Khvostantsev,~L.~G.; Slesarev,~V.~N.; Brazhkin,~V.~V. Toroid type high-pressure
  device: history and prospects. \emph{High Pressure Research} \textbf{2004},
  \emph{24}, 371--383\relax
\mciteBstWouldAddEndPuncttrue
\mciteSetBstMidEndSepPunct{\mcitedefaultmidpunct}
{\mcitedefaultendpunct}{\mcitedefaultseppunct}\relax
\EndOfBibitem
\bibitem[Fang \latin{et~al.}(2012)Fang, Bull, Loveday, Nelmes, and
  Kamenev]{Fang2012_DT_Anvils}
Fang,~J.; Bull,~C.~L.; Loveday,~J.~S.; Nelmes,~R.~J.; Kamenev,~K.~V. {Strength
  Analysis and Optimisation of Double-Toroidal Anvils for High-Pressure
  Research}. \emph{Rev. Sci. Inst.} \textbf{2012}, \emph{83}, 093902, DOI:
  \doi{10.1063/1.4746993}\relax
\mciteBstWouldAddEndPuncttrue
\mciteSetBstMidEndSepPunct{\mcitedefaultmidpunct}
{\mcitedefaultendpunct}{\mcitedefaultseppunct}\relax
\EndOfBibitem
\bibitem[Coelho(2018)]{coelho2018topas}
Coelho,~A.~A. TOPAS and TOPAS-Academic: an optimization program integrating
  computer algebra and crystallographic objects written in C++. \emph{J. Appl.
  Cryst.} \textbf{2018}, \emph{51}, 210--218\relax
\mciteBstWouldAddEndPuncttrue
\mciteSetBstMidEndSepPunct{\mcitedefaultmidpunct}
{\mcitedefaultendpunct}{\mcitedefaultseppunct}\relax
\EndOfBibitem
\bibitem[Hohenberg and Kohn(1964)Hohenberg, and Kohn]{DFT-HK}
Hohenberg,~P.; Kohn,~W. Inhomogeneous electron gas. \emph{Phys. Rev.}
  \textbf{1964}, \emph{136}, B864--B871\relax
\mciteBstWouldAddEndPuncttrue
\mciteSetBstMidEndSepPunct{\mcitedefaultmidpunct}
{\mcitedefaultendpunct}{\mcitedefaultseppunct}\relax
\EndOfBibitem
\bibitem[Kohn and Sham(1965)Kohn, and Sham]{DFT-KS}
Kohn,~W.; Sham,~L.~J. Self-consistent equations including exchange and
  correlation effects. \emph{Phys. Rev.} \textbf{1965}, \emph{140},
  A1133--A1138\relax
\mciteBstWouldAddEndPuncttrue
\mciteSetBstMidEndSepPunct{\mcitedefaultmidpunct}
{\mcitedefaultendpunct}{\mcitedefaultseppunct}\relax
\EndOfBibitem
\bibitem[Sun \latin{et~al.}(2015)Sun, Ruzsinszky, and Perdew]{SCAN-DFT}
Sun,~J.; Ruzsinszky,~A.; Perdew,~J.~P. {Strongly Constrained and Appropriately
  Normed Semilocal Density Functional}. \emph{Phys. Rev. Lett.} \textbf{2015},
  \emph{115}, 036402, DOI: \doi{10.1103/PhysRevLett.115.036402}\relax
\mciteBstWouldAddEndPuncttrue
\mciteSetBstMidEndSepPunct{\mcitedefaultmidpunct}
{\mcitedefaultendpunct}{\mcitedefaultseppunct}\relax
\EndOfBibitem
\bibitem[Dudarev \latin{et~al.}(1998)Dudarev, Botton, Savrasov, Humphreys, and
  Sutton]{Dudarev_LSDAU_1998}
Dudarev,~S.~L.; Botton,~G.~A.; Savrasov,~S.~Y.; Humphreys,~C.~J.; Sutton,~A.~P.
  {Electron-Energy-Loss Spectra and the Structural Stability of Nickel Oxide:
  An LSDA+U Study}. \emph{Phys. Rev. B} \textbf{1998}, \emph{57}, 1505--1509,
  DOI: \doi{10.1103/PhysRevB.57.1505}\relax
\mciteBstWouldAddEndPuncttrue
\mciteSetBstMidEndSepPunct{\mcitedefaultmidpunct}
{\mcitedefaultendpunct}{\mcitedefaultseppunct}\relax
\EndOfBibitem
\bibitem[Wang \latin{et~al.}(2006)Wang, Maxisch, and Ceder]{Wang_TMO+U_2006}
Wang,~L.; Maxisch,~T.; Ceder,~G. {Oxidation Energies of Transition Metal Oxides
  Within the $\mathrm{GGA}+\mathrm{U}$ Framework}. \emph{Phys. Rev. B}
  \textbf{2006}, \emph{73}, 195107, DOI: \doi{10.1103/PhysRevB.73.195107}\relax
\mciteBstWouldAddEndPuncttrue
\mciteSetBstMidEndSepPunct{\mcitedefaultmidpunct}
{\mcitedefaultendpunct}{\mcitedefaultseppunct}\relax
\EndOfBibitem
\bibitem[Monkhorst and Pack(1976)Monkhorst, and Pack]{MPgrid}
Monkhorst,~H.~J.; Pack,~J.~D. Special points for {B}rillouin-zone integrations.
  \emph{Phys. Rev. B} \textbf{1976}, \emph{13}, 5188--5192\relax
\mciteBstWouldAddEndPuncttrue
\mciteSetBstMidEndSepPunct{\mcitedefaultmidpunct}
{\mcitedefaultendpunct}{\mcitedefaultseppunct}\relax
\EndOfBibitem
\bibitem[Thom \latin{et~al.}(2009)Thom, Sundstrom, and Head-Gordon]{LOBA}
Thom,~A. J.~W.; Sundstrom,~E.~J.; Head-Gordon,~M. LOBA: a localized orbital
  bonding analysis to calculate oxidation states{,} with application to a model
  water oxidation catalyst. \emph{Phys. Chem. Chem. Phys.} \textbf{2009},
  \emph{11}, 11297--11304, DOI: \doi{10.1039/B915364K}\relax
\mciteBstWouldAddEndPuncttrue
\mciteSetBstMidEndSepPunct{\mcitedefaultmidpunct}
{\mcitedefaultendpunct}{\mcitedefaultseppunct}\relax
\EndOfBibitem
\bibitem[Thom \latin{et~al.}(2009)Thom, Sundstrom, and Head-Gordon]{B915364K}
Thom,~A. J.~W.; Sundstrom,~E.~J.; Head-Gordon,~M. LOBA: a localized orbital
  bonding analysis to calculate oxidation states{,} with application to a model
  water oxidation catalyst. \emph{Phys. Chem. Chem. Phys.} \textbf{2009},
  \emph{11}, 11297--11304, DOI: \doi{10.1039/B915364K}\relax
\mciteBstWouldAddEndPuncttrue
\mciteSetBstMidEndSepPunct{\mcitedefaultmidpunct}
{\mcitedefaultendpunct}{\mcitedefaultseppunct}\relax
\EndOfBibitem
\bibitem[Shao \latin{et~al.}(2015)Shao, Gan, Epifanovsky, Gilbert, Wormit,
  Kussmann, Lange, Behn, Deng, Feng, Ghosh, Goldey, Horn, Jacobson, Kaliman,
  Khaliullin, Ku?<9B>, Landau, Liu, Proynov, Rhee, Richard, Rohrdanz, Steele,
  Sundstrom, III, Zimmerman, Zuev, Albrecht, Alguire, Austin, Beran, Bernard,
  Berquist, Brandhorst, Bravaya, Brown, Casanova, Chang, Chen, Chien, Closser,
  Crittenden, Diedenhofen, Jr., Do, Dutoi, Edgar, Fatehi, Fusti-Molnar,
  Ghysels, Golubeva-Zadorozhnaya, Gomes, Hanson-Heine, Harbach, Hauser,
  Hohenstein, Holden, Jagau, Ji, Kaduk, Khistyaev, Kim, Kim, King, Klunzinger,
  Kosenkov, Kowalczyk, Krauter, Lao, Laurent, Lawler, nd~Ching Yeh~Lin, Liu,
  Livshits, Lochan, Luenser, Manohar, Manzer, Mao, Mardirossian, Marenich,
  Maurer, Mayhall, Neuscamman, Oana, Olivares-Amaya, O?<80><99>Neill, Parkhill,
  Perrine, Peverati, Prociuk, Rehn, Rosta, Russ, Sharada, Sharma, Small, Sodt,
  Stein, Stück, Su, Thom, Tsuchimochi, Vanovschi, Vogt, Vydrov, Wang, Watson,
  Wenzel, White, Williams, Yang, Yeganeh, Yost, You, Zhang, Zhang, Zhao,
  Brooks, Chan, Chipman, Cramer, III, Gordon, Hehre, Klamt, III, Schmidt,
  Sherrill, Truhlar, Warshel, Xu, Aspuru-Guzik, Baer, Bell, Besley, Chai,
  Dreuw, Dunietz, Furlani, Gwaltney, Hsu, Jung, Kong, Lambrecht, Liang,
  Ochsenfeld, Rassolov, Slipchenko, Subotnik, Voorhis, Herbert, Krylov, Gill,
  and Head-Gordon]{doi:10.1080/00268976.2014.952696}
Shao,~Y. \latin{et~al.}  {Advances in Molecular Quantum Chemistry Contained in
  the Q-Chem 4 Program Package}. \emph{Molecular Physics} \textbf{2015},
  \emph{113}, 184--215, DOI: \doi{10.1080/00268976.2014.952696}\relax
\mciteBstWouldAddEndPuncttrue
\mciteSetBstMidEndSepPunct{\mcitedefaultmidpunct}
{\mcitedefaultendpunct}{\mcitedefaultseppunct}\relax
\EndOfBibitem
\bibitem[Perdew \latin{et~al.}(1996)Perdew, Burke, and Ernzerhof]{PBE}
Perdew,~J.~P.; Burke,~K.; Ernzerhof,~M. {Generalized Gradient Approximation
  Made Simple}. \emph{Phys. Rev. Lett.} \textbf{1996}, \emph{77}, 3865--3868,
  DOI: \doi{10.1103/PhysRevLett.77.3865}\relax
\mciteBstWouldAddEndPuncttrue
\mciteSetBstMidEndSepPunct{\mcitedefaultmidpunct}
{\mcitedefaultendpunct}{\mcitedefaultseppunct}\relax
\EndOfBibitem
\bibitem[Weigend and Ahlrichs(2005)Weigend, and Ahlrichs]{B508541A}
Weigend,~F.; Ahlrichs,~R. Balanced basis sets of split valence{,} triple zeta
  valence and quadruple zeta valence quality for H to Rn: Design and assessment
  of accuracy. \emph{Phys. Chem. Chem. Phys.} \textbf{2005}, \emph{7},
  3297--3305, DOI: \doi{10.1039/B508541A}\relax
\mciteBstWouldAddEndPuncttrue
\mciteSetBstMidEndSepPunct{\mcitedefaultmidpunct}
{\mcitedefaultendpunct}{\mcitedefaultseppunct}\relax
\EndOfBibitem
\bibitem[Momma and Izumi(2011)Momma, and Izumi]{momma2011vesta}
Momma,~K.; Izumi,~F. VESTA 3 for three-dimensional visualization of crystal,
  volumetric and morphology data. \emph{J. Appl. Cryst.} \textbf{2011},
  \emph{44}, 1272--1276\relax
\mciteBstWouldAddEndPuncttrue
\mciteSetBstMidEndSepPunct{\mcitedefaultmidpunct}
{\mcitedefaultendpunct}{\mcitedefaultseppunct}\relax
\EndOfBibitem
\bibitem[Sales \latin{et~al.}(1997)Sales, Mandrus, Chakoumakos, Keppens, and
  Thompson]{sales1997filled}
Sales,~B.; Mandrus,~D.; Chakoumakos,~B.; Keppens,~V.; Thompson,~J. Filled
  skutterudite antimonides: Electron crystals and phonon glasses. \emph{Phys.
  Rev. B} \textbf{1997}, \emph{56}, 15081\relax
\mciteBstWouldAddEndPuncttrue
\mciteSetBstMidEndSepPunct{\mcitedefaultmidpunct}
{\mcitedefaultendpunct}{\mcitedefaultseppunct}\relax
\EndOfBibitem
\bibitem[Aktas \latin{et~al.}(2011)Aktas, Truong, Otani, Balakrishnan, Clouter,
  Kimura, and Quirion]{Aktas_2011}
Aktas,~O.; Truong,~K.~D.; Otani,~T.; Balakrishnan,~G.; Clouter,~M.~J.;
  Kimura,~T.; Quirion,~G. {Raman Scattering Study of Delafossite
  Magnetoelectric Multiferroic Compounds: CuFeO$_2$ and CuCrO$_2$}. \emph{J.
  Phys. Cond. Matt.} \textbf{2011}, \emph{24}, 036003, DOI:
  \doi{10.1088/0953-8984/24/3/036003}\relax
\mciteBstWouldAddEndPuncttrue
\mciteSetBstMidEndSepPunct{\mcitedefaultmidpunct}
{\mcitedefaultendpunct}{\mcitedefaultseppunct}\relax
\EndOfBibitem
\bibitem[Pellicer-Porres \latin{et~al.}(2005)Pellicer-Porres, Segura,
  Mart\'{\i}nez, Saitta, Polian, Chervin, and Canny]{Ramancuga}
Pellicer-Porres,~J.; Segura,~A.; Mart\'{\i}nez,~E.; Saitta,~A.~M.; Polian,~A.;
  Chervin,~J.~C.; Canny,~B. {Vibrational Properties of Delafossite
  $\mathrm{Cu}\mathrm{Ga}{\mathrm{O}}_{2}$ at Ambient and High Pressures}.
  \emph{Phys. Rev. B} \textbf{2005}, \emph{72}, 064301, DOI:
  \doi{10.1103/PhysRevB.72.064301}\relax
\mciteBstWouldAddEndPuncttrue
\mciteSetBstMidEndSepPunct{\mcitedefaultmidpunct}
{\mcitedefaultendpunct}{\mcitedefaultseppunct}\relax
\EndOfBibitem
\bibitem[Pellicer-Porres \latin{et~al.}(2006)Pellicer-Porres,
  Mart\'{\i}nez-Garc\'{\i}a, Segura, Rodr\'{\i}guez-Hern\'andez, Mu\~noz,
  Chervin, Garro, and Kim]{Ramancual}
Pellicer-Porres,~J.; Mart\'{\i}nez-Garc\'{\i}a,~D.; Segura,~A.;
  Rodr\'{\i}guez-Hern\'andez,~P.; Mu\~noz,~A.; Chervin,~J.~C.; Garro,~N.;
  Kim,~D. {Pressure and Temperature Dependence of the Lattice Dynamics of
  $\mathrm{Cu}\mathrm{Al}{\mathrm{O}}_{2}$ Investigated by Raman Scattering
  Experiments and Ab Initio Calculations}. \emph{Phys. Rev. B} \textbf{2006},
  \emph{74}, 184301, DOI: \doi{10.1103/PhysRevB.74.184301}\relax
\mciteBstWouldAddEndPuncttrue
\mciteSetBstMidEndSepPunct{\mcitedefaultmidpunct}
{\mcitedefaultendpunct}{\mcitedefaultseppunct}\relax
\EndOfBibitem
\bibitem[Chen \latin{et~al.}(2015)Chen, Lin, and Lee]{chen2015crednerite}
Chen,~H.-Y.; Lin,~Y.-C.; Lee,~J.-S. Crednerite-CuMnO$_2$ thin films prepared
  using atmospheric pressure plasma annealing. \emph{Appl. Surf. Sci,}
  \textbf{2015}, \emph{338}, 113--119\relax
\mciteBstWouldAddEndPuncttrue
\mciteSetBstMidEndSepPunct{\mcitedefaultmidpunct}
{\mcitedefaultendpunct}{\mcitedefaultseppunct}\relax
\EndOfBibitem
\bibitem[Stinton and Evans(2007)Stinton, and Evans]{Stinton}
Stinton,~G.~W.; Evans,~J. S.~O. {Parametric Rietveld refinement}. \emph{J.
  Appl. Cryst.} \textbf{2007}, \emph{40}, 87--95, DOI:
  \doi{10.1107/S0021889806043275}\relax
\mciteBstWouldAddEndPuncttrue
\mciteSetBstMidEndSepPunct{\mcitedefaultmidpunct}
{\mcitedefaultendpunct}{\mcitedefaultseppunct}\relax
\EndOfBibitem
\bibitem[Pawley(1981)]{pawley1981unit}
Pawley,~G. Unit-cell refinement from powder diffraction scans. \emph{J. Appl.
  Cryst.} \textbf{1981}, \emph{14}, 357--361\relax
\mciteBstWouldAddEndPuncttrue
\mciteSetBstMidEndSepPunct{\mcitedefaultmidpunct}
{\mcitedefaultendpunct}{\mcitedefaultseppunct}\relax
\EndOfBibitem
\bibitem[Kimber(2012)]{Kimber_2012}
Kimber,~S. A.~J. {Charge and Orbital Order in Frustrated Pb$_3$Mn$_7$O$_15$}.
  \emph{J. Phys. Cond. Matt.} \textbf{2012}, \emph{24}, 186002, DOI:
  \doi{10.1088/0953-8984/24/18/186002}\relax
\mciteBstWouldAddEndPuncttrue
\mciteSetBstMidEndSepPunct{\mcitedefaultmidpunct}
{\mcitedefaultendpunct}{\mcitedefaultseppunct}\relax
\EndOfBibitem
\bibitem[Jia \latin{et~al.}(2011)Jia, Zhang, Zhang, Guo, Zeng, and
  Lin]{Jia_MagFrust_2011}
Jia,~T.; Zhang,~G.; Zhang,~X.; Guo,~Y.; Zeng,~Z.; Lin,~H.~Q. {Magnetic
  Frustration in NaMnO$_2$ and CuMnO$_2$}. \emph{J. Appl. Phys. }
  \textbf{2011}, \emph{109}, 07E102, DOI: \doi{10.1063/1.3536533}\relax
\mciteBstWouldAddEndPuncttrue
\mciteSetBstMidEndSepPunct{\mcitedefaultmidpunct}
{\mcitedefaultendpunct}{\mcitedefaultseppunct}\relax
\EndOfBibitem
\bibitem[Kroumova \latin{et~al.}(2003)Kroumova, Aroyo, Perez-Mato, Kirov,
  Capillas, Ivantchev, and Wondratschek]{Bilbao_SAM}
Kroumova,~E.; Aroyo,~M.; Perez-Mato,~J.; Kirov,~A.; Capillas,~C.;
  Ivantchev,~S.; Wondratschek,~H. {Bilbao Crystallographic Server : Useful
  Databases and Tools for Phase-Transition Studies}. \emph{Phase Transitions}
  \textbf{2003}, \emph{76}, 155--170, DOI:
  \doi{10.1080/0141159031000076110}\relax
\mciteBstWouldAddEndPuncttrue
\mciteSetBstMidEndSepPunct{\mcitedefaultmidpunct}
{\mcitedefaultendpunct}{\mcitedefaultseppunct}\relax
\EndOfBibitem
\bibitem[Miller \latin{et~al.}(2009)Miller, Smith, Mackenzie, and
  Evans]{Miller2009}
Miller,~W.; Smith,~C.~W.; Mackenzie,~D.~S.; Evans,~K.~E. {Negative Thermal
  Expansion: a Review}. \emph{Journal of Materials Science} \textbf{2009},
  \emph{44}, 5441--5451, DOI: \doi{10.1007/s10853-009-3692-4}\relax
\mciteBstWouldAddEndPuncttrue
\mciteSetBstMidEndSepPunct{\mcitedefaultmidpunct}
{\mcitedefaultendpunct}{\mcitedefaultseppunct}\relax
\EndOfBibitem
\bibitem[Liu \latin{et~al.}(2018)Liu, Gao, Chen, Deng, Lin, and
  Xing]{C8CC01153B}
Liu,~Z.; Gao,~Q.; Chen,~J.; Deng,~J.; Lin,~K.; Xing,~X. Negative thermal
  expansion in molecular materials. \emph{Chem. Commun.} \textbf{2018},
  \emph{54}, 5164--5176, DOI: \doi{10.1039/C8CC01153B}\relax
\mciteBstWouldAddEndPuncttrue
\mciteSetBstMidEndSepPunct{\mcitedefaultmidpunct}
{\mcitedefaultendpunct}{\mcitedefaultseppunct}\relax
\EndOfBibitem
\bibitem[Mast \latin{et~al.}(2019)Mast, Lawler, Childs, Czerwinski,
  Sattelberger, Poineau, and Forster]{Mast2019}
Mast,~D.~S.; Lawler,~K.~V.; Childs,~B.~C.; Czerwinski,~K.~R.;
  Sattelberger,~A.~P.; Poineau,~F.; Forster,~P.~M. {An Atomistic Understanding
  of the Unusual Thermal Behavior of the Molecular Oxide Tc$_2$O$_7$}.
  \emph{Inorganic Chemistry} \textbf{2019}, \emph{58}, 5468--5475, DOI:
  \doi{10.1021/acs.inorgchem.8b02368}\relax
\mciteBstWouldAddEndPuncttrue
\mciteSetBstMidEndSepPunct{\mcitedefaultmidpunct}
{\mcitedefaultendpunct}{\mcitedefaultseppunct}\relax
\EndOfBibitem
\bibitem[Lawler \latin{et~al.}(2017)Lawler, Childs, Mast, Czerwinski,
  Sattelberger, Poineau, and Forster]{Lawler2017}
Lawler,~K.~V.; Childs,~B.~C.; Mast,~D.~S.; Czerwinski,~K.~R.;
  Sattelberger,~A.~P.; Poineau,~F.; Forster,~P.~M. {Molecular and Electronic
  Structures of M 2 O 7 (M = Mn, Tc, Re)}. \emph{Inorganic Chemistry}
  \textbf{2017}, \emph{56}, 2448--2458, DOI:
  \doi{10.1021/acs.inorgchem.6b02503}\relax
\mciteBstWouldAddEndPuncttrue
\mciteSetBstMidEndSepPunct{\mcitedefaultmidpunct}
{\mcitedefaultendpunct}{\mcitedefaultseppunct}\relax
\EndOfBibitem
\bibitem[Abramchuk \latin{et~al.}(2018)Abramchuk, Lebedev, Hellman, Bahrami,
  Mordvinova, Krizan, Metz, Broido, and Tafti]{abramchuk2018crystal}
Abramchuk,~M.; Lebedev,~O.~I.; Hellman,~O.; Bahrami,~F.; Mordvinova,~N.~E.;
  Krizan,~J.~W.; Metz,~K.~R.; Broido,~D.; Tafti,~F. Crystal Chemistry and
  Phonon Heat Capacity in Quaternary Honeycomb Delafossites:
  Cu[Li1/3Sn2/3]O$_2$ and Cu[Na1/3Sn2/3]O$_2$. \emph{Inorganic Chemistry}
  \textbf{2018}, \emph{57}, 12709--12717\relax
\mciteBstWouldAddEndPuncttrue
\mciteSetBstMidEndSepPunct{\mcitedefaultmidpunct}
{\mcitedefaultendpunct}{\mcitedefaultseppunct}\relax
\EndOfBibitem
\bibitem[Garg \latin{et~al.}(2014)Garg, Mishra, Pandey, and
  Sharma]{garg2014multiferroic}
Garg,~A.~B.; Mishra,~A.; Pandey,~K.; Sharma,~S.~M. Multiferroic CuCrO$_2$ under
  high pressure: In situ X-ray diffraction and Raman spectroscopic studies.
  \emph{J. Appl. Phys. } \textbf{2014}, \emph{116}, 133514\relax
\mciteBstWouldAddEndPuncttrue
\mciteSetBstMidEndSepPunct{\mcitedefaultmidpunct}
{\mcitedefaultendpunct}{\mcitedefaultseppunct}\relax
\EndOfBibitem
\bibitem[Busing and Levy(1964)Busing, and Levy]{busing1964effect}
Busing,~W.; Levy,~H. The effect of thermal motion on the estimation of bond
  lengths from diffraction measurements. \emph{Acta Crystallographica}
  \textbf{1964}, \emph{17}, 142--146\relax
\mciteBstWouldAddEndPuncttrue
\mciteSetBstMidEndSepPunct{\mcitedefaultmidpunct}
{\mcitedefaultendpunct}{\mcitedefaultseppunct}\relax
\EndOfBibitem
\bibitem[Chapman \latin{et~al.}(2005)Chapman, Chupas, and
  Kepert]{chapman2005direct}
Chapman,~K.~W.; Chupas,~P.~J.; Kepert,~C.~J. Direct observation of a transverse
  vibrational mechanism for negative thermal expansion in Zn(CN)$_2$: an atomic
  pair distribution function analysis. \emph{Journal of the American Chemical
  Society} \textbf{2005}, \emph{127}, 15630--15636\relax
\mciteBstWouldAddEndPuncttrue
\mciteSetBstMidEndSepPunct{\mcitedefaultmidpunct}
{\mcitedefaultendpunct}{\mcitedefaultseppunct}\relax
\EndOfBibitem
\bibitem[Chapman and Chupas(2009)Chapman, and Chupas]{chapman2009anomalous}
Chapman,~K.~W.; Chupas,~P.~J. Anomalous thermal expansion of cuprites: A
  combined high resolution pair distribution function and geometric analysis.
  \emph{Chem. Mater,} \textbf{2009}, \emph{21}, 425--431\relax
\mciteBstWouldAddEndPuncttrue
\mciteSetBstMidEndSepPunct{\mcitedefaultmidpunct}
{\mcitedefaultendpunct}{\mcitedefaultseppunct}\relax
\EndOfBibitem
\bibitem[Levy \latin{et~al.}(2020)Levy, Greenberg, Layek, Pasternak, Kantor,
  Pascarelli, Marini, Konopkova, and Rozenberg]{levy2020high}
Levy,~D.; Greenberg,~E.; Layek,~S.; Pasternak,~M.; Kantor,~I.; Pascarelli,~S.;
  Marini,~C.; Konopkova,~Z.; Rozenberg,~G.~K. High-pressure structural and
  electronic properties of CuMO$_2$ (M= Cr, Mn) delafossite-type oxides.
  \emph{Phys. Rev. B} \textbf{2020}, \emph{101}, 245121\relax
\mciteBstWouldAddEndPuncttrue
\mciteSetBstMidEndSepPunct{\mcitedefaultmidpunct}
{\mcitedefaultendpunct}{\mcitedefaultseppunct}\relax
\EndOfBibitem
\bibitem[Keren and Gardner(2001)Keren, and Gardner]{keren2001frustration}
Keren,~A.; Gardner,~J.~S. Frustration driven lattice distortion: An NMR
  investigation of Y$_2$Mo$_2$O$_7$. \emph{Phys. Rev. Lett.} \textbf{2001},
  \emph{87}, 177201\relax
\mciteBstWouldAddEndPuncttrue
\mciteSetBstMidEndSepPunct{\mcitedefaultmidpunct}
{\mcitedefaultendpunct}{\mcitedefaultseppunct}\relax
\EndOfBibitem
\bibitem[Kimber \latin{et~al.}(2014)Kimber, Mazin, Shen, Jeschke, Streltsov,
  Argyriou, Valenti, and Khomskii]{kimber2014valence}
Kimber,~S.~A.; Mazin,~I.; Shen,~J.; Jeschke,~H.~O.; Streltsov,~S.~V.;
  Argyriou,~D.~N.; Valenti,~R.; Khomskii,~D.~I. Valence bond liquid phase in
  the honeycomb lattice material Li$_2$RuO$_3$. \emph{Phys. Rev. B}
  \textbf{2014}, \emph{89}, 081408\relax
\mciteBstWouldAddEndPuncttrue
\mciteSetBstMidEndSepPunct{\mcitedefaultmidpunct}
{\mcitedefaultendpunct}{\mcitedefaultseppunct}\relax
\EndOfBibitem
\bibitem[Perversi \latin{et~al.}(2019)Perversi, Pachoud, Cumby, Hudspeth,
  Wright, Kimber, and Attfield]{perversi2019co}
Perversi,~G.; Pachoud,~E.; Cumby,~J.; Hudspeth,~J.~M.; Wright,~J.~P.;
  Kimber,~S.~A.; Attfield,~J.~P. Co-emergence of magnetic order and structural
  fluctuations in magnetite. \emph{Nat. Commun.} \textbf{2019}, \emph{10},
  1--6\relax
\mciteBstWouldAddEndPuncttrue
\mciteSetBstMidEndSepPunct{\mcitedefaultmidpunct}
{\mcitedefaultendpunct}{\mcitedefaultseppunct}\relax
\EndOfBibitem
\end{mcitethebibliography}

\end{document}